\documentclass[12pt,preprint]{aastex}
\shorttitle{SN~2005hk}
\shortauthors{Phillips et al.}

\received{2006 July 1}
\begin{document}

\title{The Peculiar SN 2005hk: Do Some Type~Ia
       Supernovae Explode as Deflagrations?\altaffilmark{1,2,3}}

\author{M.~M.~Phillips\altaffilmark{4},
Weidong~Li\altaffilmark{5},
Joshua~A.~Frieman\altaffilmark{6,7,8},
S.~I.~Blinnikov\altaffilmark{9,10},
Darren~DePoy\altaffilmark{11},
Jos\'e~L.~Prieto\altaffilmark{11},
P.~Milne\altaffilmark{12},
Carlos Contreras\altaffilmark{4},
Gast\'on Folatelli\altaffilmark{4},
Nidia Morrell\altaffilmark{4},
Mario Hamuy\altaffilmark{13},
Nicholas~B.~Suntzeff\altaffilmark{14},
Miguel Roth\altaffilmark{4},
Sergio Gonz\'alez\altaffilmark{4},
Wojtek Krzeminski\altaffilmark{4},
Alexei~V.~Filippenko\altaffilmark{5},
Wendy~L.~Freedman\altaffilmark{15},
Ryan~Chornock\altaffilmark{5},
Saurabh~Jha\altaffilmark{5,16},
Barry~F.~Madore\altaffilmark{15,17},
S.~E.~Persson\altaffilmark{15},
Christopher~R.~Burns\altaffilmark{15},
Pamela~Wyatt\altaffilmark{15},
David~Murphy\altaffilmark{15},
Ryan~J.~Foley\altaffilmark{5},
Mohan Ganeshalingam\altaffilmark{5},
Franklin~J.~D.~Serduke\altaffilmark{5},
Kevin~Krisciunas,\altaffilmark{18},
Bruce~Bassett\altaffilmark{19,20},
Andrew~Becker\altaffilmark{21},
Ben~Dilday\altaffilmark{7,22},
J.~Eastman\altaffilmark{11},
Peter~M.~Garnavich\altaffilmark{18},
Jon~Holtzman\altaffilmark{23},
Richard~Kessler\altaffilmark{7,24},
Hubert~Lampeitl\altaffilmark{25},
John~Marriner\altaffilmark{8},
S.~Frank\altaffilmark{11},
J.~L.~Marshall\altaffilmark{11},
Gajus~Miknaitis\altaffilmark{8},
Masao~Sako\altaffilmark{26},
Donald~P.~Schneider\altaffilmark{27},
Kurt~van~der~Heyden\altaffilmark{19},
and Naoki~Yasuda\altaffilmark{28}}

\altaffiltext{1}{Based in part on observations taken at the Cerro Tololo
Inter-American Observatory, National Optical Astronomy Observatory, 
which is operated by the Association of Universities for Research in 
Astronomy, Inc. (AURA) under cooperative agreement with the National 
Science Foundation.}
\altaffiltext{2}{Based in part on observations obtained with the Apache
Point Observatory 3.5-meter telescope, which is owned and operated
by the Astrophysical Research Consortium.}
\altaffiltext{3}{Partly based on observations collected at the European 
Southern Observatory, Chile, in the course of programme 076.A-0156}

\altaffiltext{4}{Las Campanas Observatory, Carnegie Observatories,
  Casilla 601, La Serena, Chile.}
\altaffiltext{5}{Department of Astronomy, University of California,
  Berkeley, CA 94720-3411.}
\altaffiltext{6}{Department of Astronomy and Astrophysics, The 
  University of Chicago, 5640 South Ellis Avenue, Chicago, IL 60637.}
\altaffiltext{7}{Kavli Institute for Cosmological Physics, The 
  University of Chicago, 5640 South Ellis Avenue Chicago, IL 60637.}
\altaffiltext{8}{Fermi National Accelerator Laboratory, P.O. Box 500, 
  Batavia, IL 60510.}
\altaffiltext{9}{ITEP, 117218 Moscow, Russia.}
\altaffiltext{10}{Max-Planck-Institut f\"{u}r Astrophysik, 
  Karl-Schwarzschild-Str. 1, 85741 Garching, Germany.}
\altaffiltext{11}{Department of Astronomy, Ohio State University, 140 
  West 18th Avenue, Columbus, OH 43210-1173.}
\altaffiltext{12}{Steward Observatory, 933 North Cherry Ave., Room N204,
  Tucson, AZ 85721.}
\altaffiltext{13}{Universidad de Chile, Departamento de Astronom\'{\i}a,
  Casilla 36-D, Santiago, Chile.}
\altaffiltext{14}{Texas A\&M University Physics Department, College
  Station, TX 77843-4242.}
\altaffiltext{15}{Observatories of the Carnegie Institution of
  Washington, 813 Santa Barbara St., Pasadena, CA 91101.}
\altaffiltext{16}{Kavli Institute for Particle Astrophysics and Cosmology,
  Stanford Linear Accelerator Center, 2575 Sand Hill Rd. MS 29, Menlo Park,
  CA 94025.}
\altaffiltext{17}{Infrared Processing and Analysis Center, Caltech/Jet
  Propulsion Laboratory, Pasadena, CA 91125.}
\altaffiltext{18}{Department of Physics and Astronomy, University of 
  Notre Dame, 225 Nieuwland Science, Notre Dame, IN 46556-5670.}
\altaffiltext{19}{South African Astronomical Observatory, Cape Town, 
  South Africa.}
\altaffiltext{20}{Applied Mathematics Department, University of Cape 
  Town, Cape Town, South Africa.}
\altaffiltext{21}{Department of Astronomy, University of Washington, 
  Box 351580, Seattle, WA 98195.}
\altaffiltext{22}{Department of Physics, The University of Chicago,
  5640 South Ellis Avenue, Chicago, IL 60637.}
\altaffiltext{23}{Department of Astronomy, New Mexico State University, 
  Dept. 4500, Las Cruces, NM  88003.}
\altaffiltext{24}{Center for Astrophysical Thermonuclear Flashes,
  University of Chicago, Chicago, IL 60637.}
\altaffiltext{25}{Space Telescope Science Institute, 3700 San Martin 
  Drive, Baltimore, MD 21218.}
\altaffiltext{26}{Department of Physics and Astronomy, University of 
  Pennsylvania, Philadelphia, PA 19104.}
\altaffiltext{27}{Department of Astronomy and Astrophysics, The 
  Pennsylvania State University, 525 Davey Laboratory, University Park, 
  PA 16802.}
\altaffiltext{28}{Institute for Cosmic Ray Research, University of Tokyo, 
  Kashiwa, 277-8582, Japan.}

\email{mmp, ccontreras, gfolatelli, nmorrell, miguel, sgonzalez, wojtek@lco.cl \\
weidong, rfoley, chornock, mganesh, alex@astro.berkeley.edu \\
saurabh@slac.stanford.edu \\
frieman, gm, marriner@fnal.gov \\
sergei.blinnikov@itep.ru \\
depoy, prieto, frank, jdeast@astronomy.ohio-state.edu \\
pmilne@as.arizona.edu \\
mhamuy@das.uchile.cl \\
suntzeff@physics.tamu.edu \\
wendy, persson, david, cburns, jmarshall, plwyatt@ociw.edu \\
barry@ipac.caltech.edu \\
Frank@Serduke.com \\
kkrisciu, peter.m.garnavich.1@nd.edu \\
bruce, heyden@saao.ac.za \\
becker@astro.washington.edu \\
bdilday@uchicago.edu \\
holz@nmsu.edu \\
kessler@hep.uchicago.edu \\
lampeitl@stsci.edu \\
masao@sas.upenn.edu \\
dps@astro.psu.edu \\
yasuda@icrr.u-tokyo.ac.jp}

\begin{abstract}
  We present extensive $u'g'r'i'BVRIYJHK_s$ photometry and optical
  spectroscopy of the Type Ia supernova (SN) 2005hk.  
These data reveal that SN~2005hk was nearly
  identical in its observed properties to SN~2002cx, which has been called 
  ``the most peculiar known Type~Ia supernova.''  Both supernovae exhibited
  high-ionization SN~1991T-like pre-maximum spectra, yet low peak 
  luminosities like SN~1991bg.  The spectra reveal that SN~2005hk, like 
  SN~2002cx, exhibited expansion velocities that were roughly half those of 
  typical Type~Ia supernovae.  The $R$ and $I$ light curves of both supernovae
  were also peculiar in not displaying the secondary maximum observed for
  normal Type~Ia supernovae.  Our $YJH$ photometry of SN~2005hk
  reveals the same peculiarity in the near-infrared.
  By combining our optical and near-infrared photometry of SN~2005hk with
  published ultraviolet light curves obtained with the Swift satellite, 
  we are able to construct a bolometric light curve from $\sim$15~days 
  before to $\sim$60~days after $B$~maximum.
  The shape and unusually low peak luminosity of this light curve, plus the
  low expansion velocities and absence of a secondary maximum at red and 
  near-infrared wavelengths, are all in reasonable agreement with
  model calculations of a three-dimensional deflagration that 
  produces $\sim$0.25 M$_{\odot}$ of $^{56}$Ni.
\end{abstract}


\section{Introduction}
\label{sec:intro}

More than 45 years ago, \citet{Hoyle60} first recognized that Type~Ia 
supernovae \citep[SNe~Ia; for a review of supernova classification, see][]
{Fil_97} were the observational signature of the thermonuclear 
disruption of a degenerate star.  Over the intervening years, progress has 
been slow in identifying the progenitor systems of these objects and 
understanding the details of the explosion mechanism.  At present, the 
most popular model for the progenitors of typical SNe~Ia is an accreting 
C/O white dwarf approaching the Chandrasekhar mass limit in a binary 
system.  As to the explosion mechanism, there is general consensus 
that pure deflagration models -- ones where the nuclear burning front 
remains subsonic throughout the entire explosion -- do not produce 
sufficient $^{56}$Ni and kinetic energy, are too mixed, and leave behind 
too much unburned carbon and oxygen to account for normal-luminosity 
SNe~Ia \citep{Gam_etal04,Bli_etal06}.  Rather, the explosion likely
begins as a subsonic flame that at some point converts to a slightly 
supersonic detonation \citep{Gam_etal04}.  Such ``delayed detonation'' models 
have been shown to reproduce the general observational characteristics of 
typical SNe~Ia \citep[e.g., see][]{Hof_etal03}, although these results 
await confirmation from detailed three-dimensional (3D) modeling.

SNe~Ia have been shown to be excellent 
cosmological standard candles that can be observed at epochs when the 
Universe was a third or less of its present age \citep[for a review, 
see][]{Fil_05}. After application
of a luminosity correction based on the decline rate from maximum
brightness (or the
light-curve width), distance measurements to a precision of 10\% or better 
are possible using SNe~Ia \citep{Ham_etal95, Ham_etal96, Rie_etal96, 
Phi_etal99, Jha02, Guy_etal05, Pri_etal06}.  
At near-infrared wavelengths, SNe~Ia are nearly perfect standard candles, 
yielding a distance precision better than 10\% without the need for any 
luminosity correction \citep{Kri_etal04a}.  SNe~Ia were responsible for
the discovery that the Universe is presently accelerating
\citep{Rie_etal98,Per_etal99}, and arguably provide the most precise
distances needed to calculate the equation-of-state parameter of the 
dark energy responsible for this acceleration.

The overwhelming majority of SNe~Ia obey the peak luminosity vs.
decline-rate relationship and display a remarkably uniform spectral evolution.
Objects such as SN~1991T and SN~1991bg, which were originally considered to be 
peculiar \citep{Fil_etal92a,Fil_etal92b,Phi_etal92,Leib_etal93}, may 
simply be examples of the high-luminosity and
low-luminosity extremes of the overall sequence of normal SNe~Ia
\citep{Nug_etal95}.  However, a handful of SNe~Ia truly stand out as
peculiar.  Among these is SN~2002cx, which was labeled by
\citet[][hereafter LFC]{Li_etal03} as ``the
most peculiar known Type~Ia supernova.''  These authors presented 
optical photometry and spectroscopy of SN~2002cx that revealed several 
strange properties including a high-ionization SN~1991T-like 
maximum-light spectrum dominated by iron-group elements, 
expansion velocities approximately half those of ordinary SNe~Ia, and the 
absence of secondary maxima in the $R$ and
$I$ bands.  Despite displaying a relatively normal initial decline rate of
$\Delta$m$_{15}$($B$)$ \approx 1.3$ mag, SN~2002cx had a peak luminosity more 
like that of the fast-declining ($\Delta$m$_{15}$($B$)$ \approx 1.9$ mag) 
SN~1991bg, yet faded only $\sim$3.5 mag in the $R$ band over the next nine months
compared to typical SNe~Ia which decline in brightness by $\sim$6 mag over 
the same period \citep{Jha_etal06}.  Optical spectra obtained $\sim$7--9 
months after maximum were also peculiar in revealing permitted P-Cygni 
emission lines of \ion{Fe}{2} and intermediate-mass elements such as Na and 
Ca at a phase when normal SN~Ia spectra are dominated by forbidden emission 
lines of Fe \citep{Jha_etal06}.

\citet{Bra_etal04} and \citet{Jha_etal06} have speculated that 
the peculiar nature of SN~2002cx may be consistent with the pure deflagration 
of a Chandrasekhar-mass white dwarf.  Alternatively, \citet{Kas_etal04}
suggested that SN~2002cx might be a subluminous SN~1991bg-like explosion
viewed through a hole in the ejecta.  Recently, four additional examples of
the SN~2002cx phenomenon have been identified: SNe~2003gq \citep{Jha_etal06}, 
2005P \citep{Jha_etal06}, 2005cc \citep{Ant_etal05}, and 2005hk 
\citep{Jha_etal06}.  All five appeared in spiral galaxies 
that exhibit clear signs of ongoing star formation, arguing
against the idea that they are SN~1991bg-like SNe viewed from a special 
angle since the latter objects occur preferentially in E and S0 galaxies 
\citep{How01,Gal_etal05}.  Hence, the SN~2002cx-like events appear to 
represent a bona fide subclass of SNe~Ia.

In this paper, we report extensive optical and near-infrared (NIR) 
observations of the SN 2002cx-like event SN~2005hk.  SN~2005hk was discovered 
\citep{Bur05} on the rise by the Lick Observatory Supernova Search (LOSS) 
with the 0.76~m Katzman Automatic Imaging Telescope 
\citep[KAIT;][]{Li_etal00,Fil_etal01} in images obtained on 
2005~Oct.~30.3 (UT dates are used throughout this paper). 
An independent discovery was made by the SDSS II Supernova Survey
\citep{Fri_etal07} on Oct.~28, with the SN not being visible in images taken 
two days earlier \citep{Bar_etal05}.
The SN was located 17$''$.2 east and 6$''$.9 north of the nucleus of 
UCG~272, an SAB(s)d: galaxy with a heliocentric recession velocity
of 3,895 km s$^{-1}$ according to the NASA/IPAC Extragalactic Database (NED).
An image showing the SN is reproduced in Figure~\ref{fig:fc}.

An initial spectrum of SN~2005hk taken by \citet{Ser_etal05} indicated
that this object was most likely a SN~1991T-like event caught 1--2 weeks before
maximum light.  However, closer analysis of this spectrum revealed unusually 
low expansion velocities (6,000--7,000 km s$^{-1}$), suggesting that SN~2005hk 
was a SN~2002cx-like object.  A spectrum obtained by the Carnegie Supernova
Project (CSP) three weeks later 
on Nov.~23.2 confirmed the close resemblance to SN~2002cx; moreover, the CSP 
$r'$ and $i'$ light curves showed the same peculiar absence of a 
secondary maximum that had distinguished SN~2002cx.  Hence, the discovery 
of SN~2005hk provided an excellent opportunity to study in detail the 
properties of this poorly understood subclass of peculiar SNe~Ia.  

This paper is the result of the pooling of independent photometric and 
spectroscopic observations carried out by the CSP, LOSS, and SDSS-II 
collaborations.
In the sections that follow, we present these different data sets and
describe in detail the optical and NIR properties of SN~2005hk.  We also
provide a definitive re-reduction of the LOSS $BVRI$ photometry of SN~2002cx,
a preliminary version of which was published by LFC.  Our NIR photometry of 
SN~2005hk is the first such data to be obtained for a SN~2002cx-like event, 
allowing us to re-construct the ultraviolet-optical-infrared (uvoir) bolometric 
light curve.  Finally, we compare our results for SN~2005hk with light 
curves and photospheric velocities derived from 3D models to test the 
hypothesis that SN~2002cx events are the observational signature of the 
pure deflagration of a Chandrasekhar-mass white dwarf.

\section{SPECTROSCOPY}
\label{sec:spec}

Seventeen spectra of SN~2005hk were obtained by our collaboration 
with various instruments and telescopes at several observatories, covering the 
supernova evolution from 8 days before the epoch of 
$B$ maximum to 67 days after.  A 
journal of these observations is provided in Table~\ref{tab:speclog} along 
with some information about the spectral characteristics.  Details of data 
acquisition and reduction procedures for the CSP spectra obtained with the 
duPont 2.5~m telescope are given by \citet{Ham06}; similar procedures were 
followed in the acquisition and reduction of the spectra in 
Table~\ref{tab:speclog} obtained with other telescopes.  Note that the Keck 
telescope spectrum obtained on 2005~Nov.~5 has been previously published
and discussed by \citet{Cho_etal06}.

Figure~\ref{fig:spmontage} shows a montage of all 17 spectra of SN~2005hk.  
The observations corresponding to the contiguous nights of 2005~Dec.~23 
and 24 ($+43$ and $+44$ days) have been combined in order to avoid crowding in 
the figure and to improve the signal-to-noise ratio. The positions of selected 
spectral features are indicated in this figure.

\section{PHOTOMETRY}
\label{sec:phot}

\subsection{CSP}
\label{sec:cspphot}

The CSP optical photometry of SN~2005hk was obtained 
with the Swope 1.0~m telescope at the Las Campanas Observatory (LCO), using
a SITe CCD and a set of SDSS $u'g'r'i'$ and
Johnson $BV$ filters. A subraster of $1200 \times 1200$ pixels was employed
which, at a scale of $0.''435$ pixel$^{-1}$, yields a
field of view of $8.'7 \times 8.'7$. Typical image quality ranged
between 1$''$ and 2$''$ full-width at half-maximum intensity (FWHM).
A photometric sequence of comparison stars in the SN field was
calibrated with the 
Swope telescope from observations of standard stars \citep{Smi_etal02, 
lan92} during five photometric
nights. Figure~\ref{fig:fc} shows the SN field and the selected comparison
stars. Tables~\ref{tab:stds1} and \ref{tab:stds2} list the 
average $u'g'r'i'$ and $BV$ magnitudes derived for these stars.
SN magnitudes in the SDSS and Johnson systems were obtained
differentially relative to the comparison stars using
point-spread-function (PSF) photometry. On every image, a PSF 
within a radius of $3''$ was fitted to the SN and to comparison stars. 
We refer readers to \citet{Ham06} for further details about the instrument 
and measurement techniques.

Although SN~2005hk was located reasonably far outside
UGC~272 (see Figure~\ref{fig:fc}), errors in the SN photometry can 
arise from poor subtraction of the underlying host-galaxy light.  
Normally this is dealt with by subtracting a template image taken when the 
SN was not present.  Such images of UGC~272 will eventually be obtained by 
the CSP when the SN has faded from visibility ($\sim$1 year after discovery).  
For the $u'g'r'i'$ bands, the lack of template images is not a serious 
problem since we can use for this purpose images of UGC~272 obtained by 
the SDSS~II Supernova Survey before the appearance of SN~2005hk.  Indeed, 
these images produce excellent galaxy subtractions as illustrated in the
top panel of Figure~\ref{fig:gBV}.  Encouraged by these results, we
attempted to use the SDSS~II images as provisional templates for the
CSP $BV$ images.  For the $B$ band, we found that the $g'$ image worked
best, whereas for $V$ an average of SDSS~II $g'$ and $r'$ images produced
the best results.  This procedure resulted in remarkably clean subtractions
as illustrated in the bottom two panels of Figure~\ref{fig:gBV}.

Comparison of photometry carried out on the template-subtracted 
$u'g'r'i'$ images with that measured without subtraction of the host galaxy 
shows that the $u'$ band is the most affected by the background.  At 40~days
after maximum, the $u'$ magnitudes measured without host-galaxy subtraction 
are systematically fainter by $\sim0.15$ mag.  In the $g'$ and $r'$ bands,
a similar but considerably smaller ($\leq$0.04 mag in $g'$ and $\leq$0.02 mag 
in $r'$) systematic error is present at the late epochs in the photometry
measured from the unsubtracted images.  In $i'$, there is no clear trend, with
the measurements indicating that errors no greater than $\sim0.02$ mag are 
present in photometry measured from the unsubtracted images.  Our 
provisional template-subtracted $BV$ photometry is fully consistent with
these results in indicating that the $B$-band data are affected more by 
the galaxy background ($\leq$0.05 mag at 40~days after maximum) than are
the $V$-band data ($\leq$0.02 mag at 40~days after maximum).

Excellent NIR photometric coverage of SN~2005hk was obtained by 
the CSP with the Swope 1.0~m telescope using a new camera called ``RetroCam.'' 
This instrument was built especially for the CSP and employs a HAWAII-1 
HgCdTe detector and a single filter wheel containing $Y$, $J$, and $H$ 
filters. Because it has no re-imaging optics, RetroCam does not operate in 
the $K_s$ band.  The scale of RetroCam is $0.537''$ pixel$^{-1}$, which is 
adequate for sampling the image quality delivered by the telescope; the
field of view is $9.'1 \times 9.'1$.
RetroCam and the CCD camera used for obtaining the CSP optical photometry 
are both mounted on a mechanical ``swivel,'' such that they can be exchanged 
with each other in a matter of minutes.  In practice, however, the CSP has
found it more effective to 
alternate between $u'g'r'i'BV$ and $YJH$ photometry every 2--3 nights on the 
Swope telescope.

A few additional epochs of $YJHK_s$ imaging of SN~2005hk were 
obtained using the Wide Field Infrared Camera (WIRC) \citep{Per_etal02}
mounted on the duPont 2.5~m telescope at LCO.  This instrument, and the 
methodology employed for using it to observe SNe, is described by 
\citet{Ham06}.  The WIRC detectors are the same type as in RetroCam.

Comparison stars in the SN field were calibrated in $YJHK_s$ using 
observations of standard stars \citep{persson98} obtained on five nights 
with RetroCam on the Swope telescope and one night with WIRC on the duPont.  
Table~\ref{tab:nirst} lists the final photometry for these stars. SN magnitudes
were computed differentially relative to the comparison stars with an
aperture of $2''$ and assuming no color terms \citep[see][for details]{Ham06}.
A $z'$-band image of the host galaxy UGC~272 obtained by SDSS~II before
the SN appeared was used as a provisional template for removing the
background galaxy light from the Swope observations, yielding surprisingly 
good subtractions\footnote{The superior sampling and seeing of the
duPont WIRC images did not allow the SDSS~II $z'$-band image of UGC~272
to be used for template subtraction of these data.  However, because of the 
better image quality, the magnitudes measured from the WIRC data
are less susceptible to host-galaxy contamination.}.  
Comparison of magnitudes measured from 
these template-subtracted $YJH$ images with those measured without 
subtraction of the host galaxy shows differences of $\sim0.01$ mag or less 
during the entire period that the SN was observed, indicating that
contamination from the background light of UGC~272 is not significant
in the NIR.

The final CSP $u'g'r'i'BV$ photometry of SN~2005hk is given in 
Table~\ref{tab:cspmags} and the $YJHK_s$ photometry in Table~\ref{tab:nir}.  
A minimum uncertainty of 0.015 mag in the optical bandpasses and 0.02 mag 
in the NIR is assumed for a single measurement based on
the typical scatter in the transformation from instrumental to standard
magnitudes of bright stars \citep{Ham06}. Figure~\ref{fig:csplc} shows
the corresponding light curves.

\subsection{KAIT}
\label{sec:kaitphot}

Broad-band $BVRI$ images of SN~2005hk were obtained using an
Apogee AP7 CCD camera with KAIT. The Apogee camera employs a 
$512 \times 512$ pixel CCD with 24 $\mu$m pixels, providing a field of view of
$6.'7 \times 6.'7$ at a scale of $0.''8$ pixel$^{-1}$.  The typical seeing
at KAIT is $\sim3''$, so the CCD images are well-sampled.

Magnitudes for SN~2005hk in the Kron-Johnson $BVRI$ system were 
measured from the KAIT data using PSF-fitting photometry.  As a general rule, 
the fitting radius of the PSF was set to the FWHM of the image, and the PSF 
radius to four times the FWHM.  Sky was taken to be the mode of an annulus 
located 20--26 pixels away from the SN.  Color terms derived from many nights
of standard-star observations were applied to these measurements.  Finally,
zeropoints of the SN magnitudes were calculated in reference to 
the sequence of local standards listed in Table~\ref{tab:stds2} and displayed 
in Figure~\ref{fig:fc}.  The local standards were calibrated via observations
of \citet{lan92} standards on several photometric nights.  The final SN
photometry is listed in Table~\ref{tab:kaitmags} and plotted in the upper panel 
of Figure~\ref{fig:kaitsdsslc}.  As no template images were available for the
KAIT observations, the SN magnitudes are probably somewhat affected by the
background light as per the discussion in \S~\ref{sec:cspphot}.

The position of SN~2005hk was observed by KAIT on Julian Date (JD) 
2453668.71 -- i.e., five days before its discovery by KAIT and three days
before the independent discovery by the SDSS II Supernova Survey.  This image,
which was obtained without a filter, yields a 3$\sigma$ upper limit of
18.91 mag for the SN as derived from artificial star experiments.

Figure~\ref{fig:cspkait} shows a comparison of the CSP and KAIT
$BV$ light curves of SN~2005hk.  While there is reasonably good agreement
($\leq$4~\%) in the measurements taken during the $\sim$20 day period centered
around the time of maximum light, at later epochs the CSP magnitudes 
in both $B$ and $V$ are systematically fainter than the KAIT values by as much
as $\sim$0.1~mag.  As illustrated in the lower two panels of 
Figure~\ref{fig:cspkait}, the differences appear to be systematically 
correlated with the magnitude of the SN.  Comparison of the CSP and KAIT 
magnitudes of the local photometric standards shows generally excellent agreement 
(see Table~\ref{tab:stds2}), and so this discrepancy must have some other 
cause.  The most likely explanations are contamination of the KAIT measurements
by host-galaxy light and/or differences in the response 
functions (atmosphere $+$ filter $+$ detector) of the CSP and KAIT $BV$ 
bandpasses.

The latter effect can be investigated through synthetic 
photometry of our spectra of SN~2005hk (see \S~\ref{sec:spec}).  Preliminary 
versions of the CSP response functions are given by \citet{Ham06}.  For
KAIT, we have constructed response functions using measurements of the 
filter throughputs, a typical detector quantum efficiency curve for the 
Apogee camera, standard aluminum reflectivity values, and an assumed 
atmospheric extinction function.  Magnitudes in $B$ and $V$ were then 
derived by convolving the response functions with spectra of SN~2005hk 
which had sufficient wavelength coverage to allow such a calculation.
Finally, the synthetic magnitudes were corrected by the same color terms 
applied to the CSP and KAIT SN photometry, and then subtracted one from 
the other (KAIT$-$CSP).  These differences, which are commonly referred to
as ``S-corrections'' \citep{Sun00, Str_etal02}, are plotted as solid circles 
joined by dashed lines in the middle two panels of Figure~\ref{fig:cspkait}.
As is seen, the S-corrections in $B$ over the first month of observations
of the SN are reasonably small, and cannot explain the
observed differences between the CSP and KAIT photometry.  In $V$, the
S-corrections are somewhat larger, evolving from $-0.04$ mag to $+0.03$ mag 
from 8 days before $B$ maximum to 24 days after.  Nevertheless, they do not
account for the observed photometric differences in this band either.  
We conclude,
therefore, that the differences between the CSP and KAIT $BV$ magnitudes
are most likely due to errors in the sky subtraction.  This hypothesis will
be checked once template images are available for both sets of data.

\subsection{CTIO}
\label{sec:ctiophot}

Images of SN~2005hk in $UBVRI$ were obtained on two nights in Nov. 2005 with 
the Cerro Tololo Inter-American 0.9~m telescope facility CCD camera.  
Magnitudes were measured via aperture photometry carried out with a $4''$
radius.  No template images were available for subtracting the background 
light of the host galaxy; from our experience with the CSP data, we can 
expect that the $UBV$ bands will be somewhat affected by background
contamination, whereas it should be close to negligible in $R$ and $I$.  
The zeropoints for the CTIO SN magnitudes were calculated with respect 
to the set of local standards listed in Table~\ref{tab:stds2} and 
displayed in Figure~\ref{fig:fc}.  These were, in turn, calibrated via 
observations of \citet{lan92} standards obtained on the same two nights.
The final SN photometry is listed in Table~\ref{tab:kaitmags} and plotted in 
the upper panel of Figure~\ref{fig:kaitsdsslc}.

\subsection{SDSS~II and MDM}
\label{sec:sdssphot}

SN~2005hk was independently discovered during the course of the 
SDSS~II Supernova Survey \citep{Fri_etal07}, which is using the SDSS Camera
and Telescope \citep{Gunn_etal98, Gunn_etal06} to image 300 square degrees
centered on the celestial equator in the Southern Galactic hemisphere. 
Ten epochs of $ugriz$
photometry were obtained between 2005 Oct.~28--Dec.~1 (U.T.). 
Details of the photometric system are given by 
\citet{Fuk_etal96} and \citet{Smi_etal02}.
Additional $griz$ imaging of SN~2005hk was obtained with the MDM 
Observatory 2.4~m telescope using a facility CCD imager \citep[RETROCAM; 
see][for a complete description of the imager]{Mor_etal05}.

Photometry of SN~2005hk on the SDSS images was carried out using the 
scene modeling code developed for SDSS~II as described by 
\citet{Hol_etal07}. A sequence of stars around the supernova was taken 
from the list of \citet{Ive_etal07}, who derived standard SDSS magnitudes 
from multiple observations taken during the main SDSS survey under 
photometric conditions.  Using these stars, frame scalings and astrometric 
solutions were derived for each
of the supernova frames, as well as for seventeen pre-supernova frames
taken as part of either the main SDSS survey or the SN survey. Finally,
the entire stack of frames was simultaneously fit for a single supernova
position, a fixed galaxy background in each filter (characterized by a grid
of galaxy intensities), and the supernova brightness in each frame.
SDSS photometry of SN~2005hk was also derived independently using an 
image-subtraction algorithm; the results are consistent with those of 
the scene modeling approach described above.

On JD 2453669.76 (one day after the KAIT non-detection), the SDSS 
observed the location of SN~2005hk, and no 
significant detection was observed in any bandpass.  Mean fluxes and 
uncertainties were measured at the location of the supernova, and upper 
limits were 
estimated by taking the magnitude corresponding to the observed flux plus 
3$\sigma$.  These yield upper limits of 24.17, 22.70, 23.14, and 22.02 for 
$griz$ using conventional (Pogson) magnitudes; the corresponding upper 
limits using the asinh magnitudes adopted by SDSS \citep{Lup_etal99} are 
24.02, 22.68, 23.04, and 21.83.

Supernova brightnesses in the MDM frames were also determined 
using the scene modeling code. For SN~2005hk, only a few reference stars 
were available in the field of view of the MDM observations, so the 
astrometric solutions and frame scalings are somewhat more uncertain. In 
addition, since the MDM observations had different response functions from 
the standard SDSS bandpasses, the photometric frame solutions included
color terms from the SDSS standard magnitudes.  To prevent
uncertainties in the frame parameters and color terms from possibly
corrupting the galaxy model (hence affecting the SDSS photometry), the MDM
data were not included in the galaxy determination, but the galaxy model
as determined from the SDSS was used (with color terms) to subtract the
galaxy from the MDM frames. The resulting SN photometry from the MDM frames
is reported on the native MDM system, since the color terms derived from
stars are likely not to apply to the spectrum of the supernova.

The final SDSS~II and MDM photometry for SN~2005hk is listed 
in Table~\ref{tab:sdssmags} and plotted in the lower panel of 
Figure~\ref{fig:kaitsdsslc}.

Figure~\ref{fig:cspsdss} displays a comparison of the CSP and 
SDSS~II $ugri$ photometry of SN~2005hk.  For each bandpass, the CSP
data have been fitted by a smooth curve, which has then been interpolated to
the epoch of the SDSS~II observations and subtracted.  In the $r$ and $i$
bands, the differences between the two data sets are small ($\leq$0.05 mag),
whereas in $u$ and $g$, there are significant systematic deviations 
which grow as large as $\sim$0.1 mag.  These differences between the CSP and
SDSS~II $gri$ photometry are fully explained in terms of differences in 
the response functions (atmosphere $+$ filter $+$ detector).  This is
illustrated in Figure~\ref{fig:cspsdss}, where S-corrections calculated 
for the $gri$ bands from our spectroscopic observations of SN~2005hk are
plotted.  The solid circles joined by the dashed lines in the difference 
plots for these bandpasses show the predicted $\Delta$mag values (SDSS$-$CSP) 
for epochs with spectroscopic coverage.  In general, these are completely
consistent with the observed differences, both in magnitude and trend.  
Unfortunately, the limited ultraviolet coverage of our spectra does not 
allow a similar comparison to be made in the $u$ band, but the deviations 
observed here are also likely to be due to differences in the bandpass
response functions.

\section{RESULTS}
\label{sec:results}

\subsection{Spectra}
\label{sec:res_spec}

Figure~\ref{fig:spmontage} shows that, before maximum brightness, the 
spectrum of SN~2005hk was dominated by
a blue continuum with only a few obvious absorption features, the strongest
of these being identified with \ion{Fe}{3}.  In the first spectrum obtained
8 days before $B$ maximum, there is no evidence for \ion{Si}{2} $\lambda$6355 
absorption, which is the identifying feature of typical SNe~Ia.  In the
spectra obtained over the next 11 days ($-7$ to +4 days), the \ion{Si}{2} 
absorption appears faintly and slowly intensifies, but never reaches a level
consistent with that in normal SNe~Ia.  By $+13$ days, the line is no longer
distinguishable.  These properties are consistent with the early spectral
evolution of SN~1991T-like events, which explains the initial report of
SN~2005hk as a likely member of this subgroup of SNe~Ia \citep{Ser_etal05}.
Nevertheless, closer examination of the early-time spectra reveals peculiarly
low expansion velocities for the \ion{Fe}{3} and \ion{Si}{2} lines 
(see below).  Moreover, the spectra obtained after maximum display unusually
sharp emission and absorption features, unlike those of typical SNe~Ia.  
\emph{These 
two characteristics are what spectroscopically distinguish SN~2005hk as a 
SN~2002cx-like object rather than an example of the SN~1991T subclass.}

The remarkable spectroscopic similarity of SN~2005hk and SN~2002cx is 
illustrated in Figure~\ref{fig:02cx_05hk_spec}.  Here spectra of SN~2005hk
at four different epochs are compared with similar-epoch spectra of
SN~2002cx taken from LFC.  The spectra are so similar that it is almost
impossible to distinguish them.  Such a close resemblance implies
that the expansion velocities of the ejecta of these two objects were also 
very similar.  This is confirmed in Figure~\ref{fig:02cx_05hk_vels}, which 
compares measurements of the expansion velocities of selected lines of 
\ion{Fe}{2}, \ion{Fe}{3}, \ion{Ca}{2}, \ion{Si}{2}, and \ion{S}{2} for both 
SNe.  These values, which were derived by estimating the wavelength of the
minimum of each 
feature, are nearly indistinguishable between the two SNe.  Moreover, as 
discussed by LFC, they are remarkably low when compared to typical expansion 
velocities of the same lines in either normal SNe~Ia or the peculiar 
SN~1991T-like objects at similar epochs.  This is illustrated in 
Figure~\ref{fig:02cx_05hk_vels}, where we have included expansion velocity
measurements of the normal SN~Ia 1992A ($\Delta$m$_{15}$($B$)$ = 1.47 \pm 
0.05$ mag).  

Following \citet{Bra_etal04}, we used the supernova synthethic spectrum
code SYNOW \citep{Fisher00} to estimate the photospheric 
velocity as the SN evolved.  Not surprisingly, considering the impressive 
similarity between the spectra of SN~2005hk and SN~2002cx, we found that 
the line identifications and SYNOW input parameters used by
\citet{Bra_etal04} to study SN~2002cx produced the best matches to the 
SN~2005hk spectra.  In general, between days $-8$ and +44, an approximately
linear decline in the photospheric velocity from $\sim7000$ km s$^{-1}$
to $\sim3000$ km s$^{-1}$ is implied.  This trend is compared in 
Figure~\ref{fig:02cx_05hk_vels} with the expansion velocities measured
from the minimum of the \ion{Si}{2} $\lambda$6355 absorption.  The
disagreement between the SYNOW predictions and the velocities deduced 
from the minima of the different lines included in 
Figure~\ref{fig:02cx_05hk_vels} illustrates the difficulty of precisely
determining the photospheric velocity from absorption minimum measurements
alone \citep[see][]{Blo_etal06}.

SYNOW is also a useful tool for checking line identifications.  
\citet{Cho_etal06} used SYNOW to analyze the 2005~Nov.~5.4 (day~$-5$) spectrum 
of SN~2005hk and found evidence for not only \ion{Fe}{3} lines, but also 
\ion{Ni}{2} and \ion{Co}{2}.  This is intriguing since one of the 
predictions of deflagration models is that material completely burned to 
the iron peak should be mixed to the outer layers of the ejecta.  Deflagration
models also leave behind significant amounts of unburned oxygen and carbon
\citep[e.g., see][]{Tra_etal04}.  \citet{Cho_etal06} found reasonably convincing 
evidence for \ion{O}{1} in the day~$-5$ spectrum of SN~2005hk, but in spite of the 
possible coincidence of features in the spectrum with the strongest lines of 
\ion{C}{2} and \ion{C}{3}, they were unable to claim a positive detection of 
carbon.  

We have used the SYNOW code to search for evidence of oxygen and 
carbon in the post-maxima spectra of SN~2005hk.  In general, we found that 
inclusion of lines of \ion{O}{1} and \ion{C}{2} did not significantly improve 
the SYNOW fits.  In particular, as previously found by \citet{Bra_etal04}, the 
strong absorption feature at $\sim$7600~\AA~which is often attributed to 
\ion{O}{1}~$\lambda$7773 can be accounted for by a blend of \ion{Fe}{2} lines.  
It may be that line blocking by \ion{Fe}{2} lines forming at similar or
higher velocities than any unburned carbon and oxygen at low velocities
may make the latter difficult to detect \citep{Bar_etal03}.
Detailed modeling of the spectrum based on the results of 3D~deflagration 
models will be required to determine if this lack of evidence 
for oxygen and carbon lines at post-maximum epochs is significant.

\subsection{Optical Light Curves and Colors}
\label{sec:res_optlc}

On the basis of the spectroscopic comparison alone, there can be little 
question that SN~2005hk and SN~2002cx are closely related.  This 
conclusion is reinforced by the photometric comparison shown in 
Figure~\ref{fig:02cx_05hk_phot}.  Here the KAIT $BVRI$ light curves of the 
two SNe have been normalized to the same brightness at peak and overplotted.
Note that the data plotted here for SN~2002cx are a re-reduction of the 
provisional photometry published by LFC.  The LFC magnitudes suffered from 
contamination by the underlying host-galaxy light due to the lack of 
appropriate template images for subtracting the galaxy.  Suitable templates 
were eventually acquired in 2005 allowing final light curves of SN~2002cx 
to be measured.  These re-reduced data are given in Appendix~A.

Figure~\ref{fig:02cx_05hk_phot} illustrates the remarkable similarity of 
the $BVRI$ light curves of SN~2002cx and SN~2005hk.  Only at epochs later 
than 15~days after $B$ maximum do there appear to be any significant 
differences, with SN~2005hk appearing to decline somewhat more slowly in $B$ 
and $V$.  K-corrections have not been applied to the photometry of either SN, 
although we can use our spectra of SN~2005hk to estimate these.  In the $R$ and 
$I$ bands, the difference in the K-corrections for SN~2002cx and SN~2005hk is 
less than $\sim$0.03~mag over the entire range of epochs from $-8$ to 
$+67$~days.  In $B$ and $V$, K(2005hk)$ - $K(2002cx) varies from 
$\sim +0.01$~mag at $-8$~days to $\sim -0.05$~mag at $+24$~days.  Beyond
this epoch, the difference in the $V$-band K-corrections slowly evolves 
back to a value of $\sim -0.02$~mag at $+67$~days. Our spectra do not have
sufficient wavelength coverage to derive the difference in the $B$-band
K-corrections beyond $+24$~days, but based on the observed color evolution,
we expect that this should also 
evolve back to values closer to 0.0~mag.  Thus, K-corrections would produce
slightly different decline rates for SN~2002cx and SN~2005hk in the same sense 
as is observed, but this effect is not sufficient to explain most of the
difference.  The photometry of SN~2005hk has not
been corrected for possible contamination from the background light of the
host galaxy.  As discussed in \S~\ref{sec:kaitphot}, there is some evidence
that the KAIT photometry of SN~2005hk suffers from such contamination, 
although this too does not appear to be sufficient to explain all of the 
late-time discrepancy seen in Figure~\ref{fig:02cx_05hk_phot}.  

An average of the KAIT and CSP $B$ light curves of SN~2005hk gives a 
decline-rate measurement of $\Delta$m$_{15}$($B$)$ = 1.56 \pm 0.09$ mag, with the 
uncertainty being dominated by the systematic photometric differences between 
the two data sets.  The decline rate of SN~2002cx must be very close to 
this number rather than the value of 1.3~mag reported by LFC\footnote{LFC 
actually suggested that the true value of $\Delta$m$_{15}$($B$) for 
SN~2002cx might be $\sim$1.6 mag after correction of the photometry for
host-galaxy contamination}.  Polynomial fits to the $B$ light curve of
SN~2002cx give
imprecise results due to the
lack of points between $-2$ and $+6$~days.  To circumvent this problem, we
used the stretch technique \citep{Gol_etal01} to fit the SN~2002cx data to
the smooth versions of the CSP $B$ and $V$ light curves of SN~2005hk shown in
Figure~\ref{fig:cspkait}.  Averaging the results for $B$ and $V$ yields
an estimate of $\Delta$m$_{15}$($B$)$ = 1.7 \pm 0.1$ mag for SN~2002cx.

The faint pre-discovery upper limits derived from the SDSS $griz$ observations
allow us to fairly precisely estimate the time of explosion of SN~2005hk.
As discussed by \citet{Rie_etal99}, during the very early rise time of a
SN~Ia, the luminosity should be approximately proportional to the square of
the time since explosion.  We therefore fit the first two detections of
SN~2005hk (JD 2453671.84 and 2453674.74) in the $griz$ bands, plus the 
3$\sigma$ upper limits on JD 2453669.76, to a second-order polynomial to
estimate the explosion date.  All four bands yielded values in the range
of $-14.8$ to $-15.4$ days before the epoch of $B$~maximum.  Hence, we conclude
that the explosion which produced SN~2005hk
most likely took place $15 \pm 1$ days before $B$~maximum.  This implies
a sharper initial rise to maximum in the $B$ band than would be expected for
a normal SN~Ia with a decline rate of $\Delta$m$_{15}$($B$)$ = 1.56$ mag.

Figure~\ref{fig:02cx_05hk_colors} compares the $B-V$, $V-R$, and
$V-I$ color evolution of SN~2005hk with that of SN~2002cx.  LFC estimated
that the host-galaxy reddening of SN~2002cx was negligible, so the data 
for this event have only been corrected for a Galactic extinction of
$E(B-V)_{\mathrm {Gal}} = 0.034$ mag.  The data for SN~2005hk are corrected 
for a Galactic reddening of $E(B-V)_{\mathrm {Gal}} = 0.022$ mag and a 
host-galaxy extinction of $E(B-V)_{\mathrm {Host}} = 0.09$ mag (see next 
paragraph). Shown for
reference is the color evolution of SN~1992A, a typical SN~Ia with a decline
rate of $\Delta$m$_{15}$($B$)$ = 1.47$ mag.  In all three colors,
the data for SN~2005hk and SN~2002cx track each other extremely 
well until $\sim$20~days after maximum.  Later than this epoch, there
is some evidence from the $B-V$ and $V-I$ colors that SN~2002cx was redder 
than SN~2005hk, although the data for SN~2002cx are of relatively poor
quality.  Except at pre-maximum epochs, the color evolution in $V-R$ and 
$V-I$ of these two SNe is considerably redder than that of the normal
SN~1992A.  As commented by LFC, in $B-V$ the color evolution 
is more similar to that of normal SNe~Ia.

For typical SNe~Ia, the color evolution can be used to estimate
the amount of host-galaxy reddening \citep{Phi_etal99}.  In the case of
SN~2005hk, such a procedure is highly problematic due to the obvious 
spectroscopic and photometric peculiarities.  Nevertheless, 
the fact that the SN was quite blue in all three colors at pre-maximum epochs 
suggests that the host-galaxy reddening is relatively small.  As noted
by \citet{Cho_etal06}, our earliest spectra of SN~2005hk reveal weak 
interstellar absorption due to the \ion{Na}{1}~D lines, both at zero redshift 
(i.e., produced by gas in our own Galaxy) and at the redshift of the host 
galaxy of the SN, UGC~272.  Dust reddening in our own Galaxy along the line
of site to SN~2005hk is estimated to be $E(B-V)_{\mathrm {Gal}} = 0.022$ mag 
\citep{Sch_etal98}, and so by analogy the extinction produced by UGC~272 is 
probably also low.  From spectropolarimetry of SN~2005hk, \citet{Cho_etal06} 
estimated a value of 0.27\% for the interstellar polarization produced by
UGC~272, which for standard dust polarization efficiencies corresponds to
$E(B-V)_{\mathrm {Host}} = 0.09$ mag.  These authors also independently estimated
the host-galaxy reddening from the interstellar \ion{Na}{1}~D lines and
found it consistent with the estimate from the polarization.  We shall 
assume this same value in this paper\footnote{If we were to assume that 
SN~2005hk and SN~2002cx had identical colors, then a relative reddening of
$E(B-I) = 0.20 \pm 0.05$ mag is implied by the photometry at times less than
20~days after the epoch of $B$ maximum.  This translates to 
$E(B-V) = 0.07 \pm 0.02$ mag, which is consistent with the estimate of the 
host-galaxy reddening of SN~2005hk derived from the polarization and 
\ion{Na}{1}~D line measurements.}.

Table~\ref{tab:appmags} lists the observed peak magnitudes in the
optical bandpasses as measured from the merged CSP, SDSS-II, and KAIT photometry.

\subsection{NIR Light Curves and Colors}
\label{sec:res_nirlc}

The NIR light curves of SN~2005hk share the same peculiar morphology 
of the $R$ and $I$ light curves in not showing a clear secondary maximum.  
Although this trait is exhibited by the fastest declining SNe~Ia, the shapes 
of the NIR light curves of such events are noticeably distinct.  This is 
illustrated in Figure~\ref{fig:nir} which compares the $YJH$ light curves of 
SN~2005hk
with those of two other SNe~Ia observed by the CSP.  The decline rates of the
latter two objects, SN~2005el and SN~2005ke, were $\Delta$m$_{15}$($B$) = 1.47
and 1.82 mag, respectively.  These values bracket the observed decline rate of
SN~2005hk of $\Delta$m$_{15}$($B$) = 1.56 mag.  As Figure~\ref{fig:nir} shows,
the $YJH$ light curves of SN~2005el display two clear peaks, as is typical
of normal SNe~Ia.  In slower-declining SNe~Ia, the second peak is more 
prominent and occurs later in time than in faster-declining events.  This
results in the secondary maximum of the fastest-declining objects like 
SN~2005ke being reduced to an inflection rather than a separate peak (see 
Figure~\ref{fig:nir}).  The morphology of the NIR light curves of SN~2005hk
is more like that of a fast-declining SN~Ia in that the secondary maximum
appears as an inflection rather than a separate peak.  However, SN~2005hk is 
very peculiar in not showing a prominent primary maximum -- rather, both the
primary and secondary maxima appear as inflections in the light curves, with
the second inflection corresponding to the actual peak of the light curve.

\citet{Kasen06} has recently emphasized that the strength
of the secondary maximum in the $IJHK$ bands is an excellent diagnostic
of the degree of $^{56}$Ni mixing in the ejecta of SNe~Ia.  In particular,
the double-peaked structure observed in the NIR light curves of typical
SNe~Ia is a direct indication of the concentration of iron-peak elements
in the central regions.  \citet{Bli_etal06} pointed out that because 
3D~deflagration models are characterized by strong mixing, they do not
do a good job of reproducing the strong secondary maxima of the $R$ and $I$
light curves of normal SNe~Ia.  The morphologies of the $YJH$ light curves of 
SN~2005hk seen in Figure~\ref{fig:nir} are strikingly similar to those of the 
theoretical $IJH$ light curves illustrated in Figure~9 of \citet{Kasen06} for a 
fiducial SN~Ia model containing 0.6 M$_{\odot}$ of $^{56}$Ni in a fully mixed 
compositional structure.  Hence, the shapes of the NIR light curves of
SN~2005hk argue in a relatively model-independent way that the ejecta
must be well-mixed.

Figure~\ref{fig:optir} compares the $V-Y$, $V-J$, and $V-H$ color 
evolution of SN~2005hk with that of SN~2005el and SN~2005ke.  Although the 
colors of all three SNe have been corrected only for Galactic reddening, the
host-galaxy reddening is probably less than or equal to $E(B-V) \approx 0.1$
mag in all three cases.  The differences between SN~2005el and SN~2005ke reflect 
their differing decline rates: SN~2005ke was a fast-declining SN~Ia, and 
therefore characterized by a lower ``effective temperature'' 
\citep{Nug_etal95} and redder colors during the initial photospheric phase 
of the explosion than was the slower-declining SN~2005el.  At phases later 
than $\sim$30 days after the time of $B$ maximum, when the spectrum had 
grown to be dominated by emission due to iron-peak elements, the color 
evolution of these two SNe was much more similar, as in the ``Lira'' effect
in $B-V$ \citep{Phi_etal99}.  The $V-$NIR color evolution of SN~2005hk is 
consistent with its peculiar nature.
Figure~\ref{fig:optir} shows that in all three colors, SN~2005hk started off
with approximately the same blue colors as SN~2005el, but by 25--30 days
after $B$ maximum had evolved to considerably redder colors than either
SN~2005el or SN~2005ke.  This behavior is broadly consistent with the 
peculiar properties of the evolution of the optical colors of SN~2005hk (see
Figure~\ref{fig:02cx_05hk_colors}).

Measurements of the apparent peak magnitudes of SN~2005hk in the NIR 
bandpasses are given in Table~\ref{tab:appmags}.  Remarkably, the date of 
maximum of the $H$ band occurred $\sim$20 days after $u'$ maximum.  In 
typical SNe~Ia, the maximum of the $H$-band light curve occurs 4--5 days
before $B$ maximum, or 1--2 days \emph{before} the epoch of maximum in $U$
\citep[e.g., see][]{Sun_etal99,Her_etal00}.

\subsection{Absolute Magnitudes and UVOIR Bolometric Light Curve}
\label{sec:res_uvoir}

To calculate the absolute magnitudes of SN~2005hk, we assume a distance 
modulus $(m-M) = 33.46 \pm 0.27$ mag for the host galaxy UGC~272 which we 
derive from the observed radial velocity in the cosmic microwave background 
frame, $v_{CMB} = 3548$ km s$^{-1}$, an assumed Hubble constant of 
$H_0 = 72$ km s$^{-1}$ Mpc$^{-1}$ \citep{Fre_etal01}, 
and a possible peculiar motion of UGC~272 of
$\pm$400 km s$^{-1}$ with respect to the Hubble flow.  Likewise, we adopt the 
values of the Galactic and host-galaxy reddenings given in the previous 
section, and assume a standard Galactic reddening law with $R_V = 3.1$ 
\citep{Car_etal89} to 
derive the extinction for any particular filter.  

With these assumptions and 
the apparent magnitudes listed in Table~\ref{tab:appmags}, we derive the
absolute $BVRIJHK$ magnitudes given in Table~\ref{tab:absmags}.  Shown for 
comparison in the same table are the absolute magnitudes of (1) SN~2002cx, 
(2) a normal SN~Ia with the same decline rate as SN~2005hk 
($\Delta$m$_{15}$($B$) = 1.56 mag), and (3) the subluminous 1991bg-like event
SN~1999by ($\Delta$m$_{15}$($B$) = 1.90 mag).  The absolute magnitudes of
SN~2002cx were calculated assuming no significant host-galaxy reddening
as per the discussion in LFC.  K-corrections derived from our spectra of
SN~2005hk were applied to the $BVRI$ magnitudes of both SN~2002cx and
SN~2005hk.  The optical magnitudes for a typical SN~Ia are 
based on the luminosity vs. decline-rate relations given in \citet{Pri_etal06}
adjusted to a Hubble constant of $H_0 = 72$ km s$^{-1}$ Mpc$^{-1}$;
the NIR values are taken from \citet{Kri_etal04b}.
Finally, the absolute magnitudes of
SN~1999by were calculated from the photometry of \citet{Gar_etal04} and
the Cepheid distance modulus of \citet{Mac_etal01}.

The numbers in Table~\ref{tab:absmags} suggest that SN~2005hk may 
have been $\sim$0.5~mag more luminous than SN~2002cx.  However, this difference
falls nearly within the errors, which are dominated by the uncertainties in 
the distances to both objects.  Table~\ref{tab:absmags} also indicates that 
both SNe were significantly subluminous compared to typical SNe~Ia.  In the 
case of SN~2005hk, the difference in absolute magnitude is $\sim$0.6--1.0~mag in 
each color except the $H$ band, where it decreases to $\sim$0.3~mag which
is within the dispersion of normal SNe~Ia.  With respect to the 
subluminous SN~1991bg-like event SN~1999by, both SN~2002cx and SN~2005hk are 
seen to be bluer, but of comparable luminosity (especially in the red).
Figure~\ref{fig:pabsmags} provides a graphical comparison of the absolute 
magnitudes of SN~2005hk with those of normal SNe~Ia in the $BVIJH$ bands,
and helps to emphasize the subluminous nature of SN~2005hk in all bands 
except $H$.  Remarkably, SNe~Ia truly do appear to be essentially 
perfect standard candles in the $H$ band with few exceptions.

We have combined the CSP $u'g'r'i'BV$ and $YJHK_s$ photometry 
of SN~2005hk with ultraviolet light curves obtained with the Swift 
\citep{Mil_etal07} satellite to compute a UVOIR bolometric light curve 
covering the wavelength range from $\lambda$ = 1100-$\infty$ \AA. 
The broad-band magnitudes were corrected for Galactic and host-galaxy 
extinction assuming a reddening law with $R_V = 3.1$ as parameterized by 
\citet{Car_etal89}.   These were then converted to monochromatic fluxes
at the effective wavelengths of each filter, and the total flux between
the Swift UVW2 band ($\lambda_{eff} = 1880$ \AA) and the $H$ band
($\lambda_{eff} = 1.63 \mu$m) was integrated using the trapezoid 
approximation.  Blueward of the blue edge of the UVW2 band at 
$\sim$1100 \AA\, we have assumed that there
is no additional flux.  To account for the missing flux redward of the
$H$ band, we have extrapolated to $\lambda = \infty$ using a blackbody (BB) 
model obtained by fitting the $g'r'i'BV$ and $YH$ fluxes with a Planck 
function.  (The $u'$ and $J$ fluxes 
were not employed in the BB fits because they appear depressed with 
respect to the other bandpasses, especially at later epochs.) 

To examine the initial rise of the UVOIR bolometric light curve, we have
incorporated photometry from the discovery images of SN~2005hk taken 
by SDSS~II on JD~2453671.84, as well as the upper limits from the pre-discovery
$griz$ images obtained two days earlier.  At these early epochs when the
SN was blue, the contribution of the NIR flux to the UVOIR light curve can 
effectively be ignored, but not the UV flux.  Since the Swift observations
did not begin until JD~2453678.3, we have calculated the relative contribution 
of the UV flux observed by Swift compared to the total UVOIR flux during the 
rise to maximum, and have extrapolated a fit to this ratio to the earlier 
epochs covered by the SDSS~II observations.  This procedure, while clearly
approximate, implies corrections to the UVOIR flux on JD~2453669.76 on 
JD~2453671.84 of 19\% and 15\%, respectively.  

The $F_{\mathrm {UVOIR}}$ values were transformed into luminosities 
assuming a distance modulus for the host galaxy of SN~2005hk, UGC~272, of 
$(m-M)_0 = 33.46$ mag (see above).  Figure~\ref{fig:uvoir} shows 
the final UVOIR bolometric light curve.  For comparison, the UVOIR light 
curve of a typical SN~Ia, 2001el ($\Delta$m$_{15}$($B$) = 1.13 mag) 
\citep{Can_etal03} is plotted.   Also included is the UVOIR light curve of the
SN~1991bg-like event SN~1999by ($\Delta$m$_{15}$($B$) = 1.90 mag), which we 
calculated from the data given by \citet{Gar_etal04}.  As discussed in 
\S~\ref{sec:res_optlc},
we have assumed for SN~2005hk that the explosion occurred 15 days before
$B$ maximum.  For SN~2001el and SN~1999by, we have taken this number to
be 18.1 and 11.0~days before $B$ maximum, respectively, which we derived from
the average $B$ light-curve rise time given by \citet{Con_etal06} after 
application of stretch values appropriate for both SNe.

Figure~\ref{fig:uvoir} emphasizes the
subluminous nature of SN~2005hk, but also shows that these events are 
\emph{not} as extreme as the SN~1991bg-like SNe~Ia.  Note 
the much slower decline rate at later epochs of SN~2005hk.  This
behavior is the opposite of what normal SNe~Ia display in that the UVOIR
bolometric light curves of lower-luminosity events decline more rapidly 
at all epochs than do higher-luminosity events \citep{Can_etal03} (cf.
SN~1999by vs. SN~2001el).  The decline rate of the bolometric light curve 
of SN~2005hk between 55--70~days after explosion was 0.027 mag day$^{-1}$,
while that for SN~2001el during the same period was 0.035 mag day$^{-1}$.
Both of these values are significantly larger than the decline rate 
of $0.0098$ mag day$^{-1}$ predicted if all the energy from the decay of 
$^{56}$Co into $^{56}$Fe were fully thermalized in the ejecta \citep{Woosley88}.

From the observed peak luminosity of SN~2005hk
of $4.3 \times 10^{49}$ erg s$^{-1}$ and application of Arnett's rule
\citep{Arn82,Arn_etal85} as per the prescription given by 
\citet{Str_etal06}, we estimate that a $^{56}$Ni mass of 
$\sim$0.22 M$_{\odot}$ was produced during the explosion of SN~2005hk
(see \S~\ref{sec:models}).
Among SNe~Ia, such a low value is only observed for the fast-declining
SN~1991bg-like events.  However, the latter objects are characterized by
lower-ionization spectra at maximum, reflecting a lower ``effective 
temperature,'' whereas SN~2005hk displayed a high-ionization, SN~1991T-like
spectrum at maximum.  Moreover, the UVOIR light curves of SN~1991bg-like events
are much narrower than that of SN~2005hk, dropping by $\sim$0.7--0.8 dex over 
the first 20 days from maximum compared to the $\sim$0.3 dex drop displayed 
by SN~2005hk (see Figure~\ref{fig:uvoir}).

\section{Comparison with 3D Deflagration Models}
\label{sec:models}

Our extensive spectroscopic and photometric observations of
SN~2005hk demonstrate conclusively that this object was nearly an exact
twin of the highly peculiar SN~Ia 2002cx.  \citet{Jha_etal06} have
identified three other SNe --- 2003gq, 2005P, and 2005cc --- which likely
also resembled SN~2002cx, and argued persuasively that SN~2002cx-like events
represent a new and distinct subclass of SNe~Ia.  The principal properties
of these objects have been summarized by \citet{Jha_etal06}; we repeat and
amplify on these below:

\begin{itemize}

\item {\em A SN~1991T-like spectrum dominated by a blue continuum and weak 
\ion{Fe}{3} absorption features at pre-maximum epochs.}  This implies that
the outermost layers of the ejecta were characterized by a hotter
``effective temperature'' than is typical of SNe~Ia of similar decline
rate \citep{Nug_etal95}.

\item {\em Iron features in the spectra at all epochs, and absence of 
secondary maxima in the light curves.}  Spectral features identified with
iron were visible in SN~2002cx from pre-maximum epochs to $\sim$9 months
after maximum with expansion velocities ranging from $\sim$7000 km~s$^{-1}$
to $\sim$700 km~s$^{-1}$ \citep{Jha_etal06}.  Secondary maxima were also
notably absent in the $IYJH$ light curves of SN~2005hk.  Both of these
observations imply that fully burned (i.e., to the iron peak) material was 
present at all layers in the ejecta.

\item {\em Features due to metals lighter than Fe at all epochs.}  
\citet{Bra_etal04}
identified lines due to Si, S, Ca, and Na in spectra of SN~2002cx covering
from pre-maximum to $\sim$2 months after maximum, and we have observed the
same features in our spectra of SN~2005hk.  \citet{Jha_etal06} unambiguously
identified lines of Ca and Na in spectra of SN~2002cx obtained $\sim$9 months
after maximum.  Hence, partially burned material is also clearly present
at all layers of the ejecta.

\item {\em Possible presence of low-velocity (500--1000 km~s$^{-1}$) 
\ion{O}{1} at late epochs.}  \citet{Jha_etal06} tentatively identified
several weak features in spectra of SN~2002cx at $\sim$9 months after
maximum with \ion{O}{1}.  If confirmed, this implies the presence of unburned
material in the inner layers.

\item {\em Low expansion velocities.}  Low expansion velocities were observed
at all epochs for SN~2002cx and SN~2005hk, implying low kinetic energy of the
ejecta.

\item {\em Low peak luminosity.}  The low peak luminosities of SN~2002cx and
SN~2005hk imply productions of $^{56}$Ni masses of $\sim$0.25 M$_{\odot}$ or
less.

\item {\em Very slowly declining UVOIR bolometric light curve at late times.}
The simplest interpretation of this property is that the ejected masses of
this subclass of SNe~Ia are relatively large, thus providing a higher opacity
to the $\gamma$-ray emission produced by radioactivity.

\item {\em Permitted \ion{Fe}{2} lines and continuum or pseudo-continuum flux
at late times.}  \citet{Jha_etal06} identified several permitted lines of
\ion{Fe}{2} with P-Cygni profiles in spectra of SN~2002cx obtained $\sim$9
months after maximum, and also noted the presence of a continuum or 
pseudo-continuum at this epoch.  These observations,
coupled with the absence of strong forbidden-line emission
other than [Ca~II] $\lambda\lambda$7291,7324, imply a relatively high
density and large mass at low velocity.

\item {\em Low level of continuum polarization.} \citet{Cho_etal06} obtained
spectropolarimetry of SN~2005hk at $-5$ days, and found a low level of 
continuum polarization ($\sim$0.4~\%) after correction for the interstellar
component.  This value is typical of normal SNe~Ia, and implies that the
peculiarities of SN~2002cx-like events cannot be explained by large
asymmetries.

\end{itemize}

\citet{Bra_etal04} and \citet{Jha_etal06} have suggested that the 
unusual properties of SN~2002cx-like events may be consistent with 3D 
deflagration models.  In general, the 3D deflagration models studied to date 
produce too little kinetic energy, too little and too mixed $^{56}$Ni, and 
too much unburned carbon and oxygen at low velocity to account for 
normal-luminosity SNe~Ia \citep{Gam_etal04,Bli_etal06}.  In addition, 
\citet{Tho_etal02} have argued that the composition structure of 3D 
deflagrations is too clumpy to produce the uniformly deep 
\ion{Si}{2} $\lambda$6355 absorption observed in typical SNe~Ia.
Interestingly, these very same failings of 3D deflagrations in explaining
normal SNe~Ia correspond closely to the observed peculiarities of
SN~2002cx-like events.  In order to pursue this idea further, we have compared 
our observations of SN~2005hk with recent calculations of synthetic bolometric 
and broad-band light curves of SNe~Ia based on four 3D deflagration 
models \citep{Bli_etal06}.

First, we select the most appropriate model of the four 
studied by \citet{Bli_etal06}
using the estimate of the $^{56}$Ni mass given in \S~\ref{sec:res_uvoir}.
Here we discuss briefly the foundation of this estimate.
\citet{Arn79,Arn82} has shown theoretically that the mass of radioactive 
$^{56}$Ni is proportional to the peak luminosity of an SN~Ia. 
Arnett's rule simply states that at the epoch of maximum light the peak 
luminosity is equal to the rate of gamma-ray deposition inside the ejecta.
The derivation of this statement is based on many simplifying assumptions,
yet the relation found by Arnett is useful for quick estimates.
A detailed description  of the $^{56}$Ni mass derivation using
broad-band optical photometry is given by \citet{Con00}.
An empirical procedure for finding the $^{56}$Ni mass based on further 
simplifications has been developed by \citet{StrLeib05}.
With a UVOIR light curve and the simple relation 
$L_{\rm max}= 2\times 10^{43} M_{\rm Ni}/M_{\sun}$
erg s$^{-1}$ (and 10\% correction taking into account the difference
of UVOIR and bolometric luminosity), they were able to make estimates 
of the  $^{56}$Ni~mass for a large number of SNe~Ia. 
The accuracy of this procedure has been tested by \citet{Bli_etal06} with 
their synthetic light curves.  It was found that for the deflagration models 
presented there, an accuracy of $\sim 20$\% may be expected.
New results by \citet{Stritz07} confirm this level of accuracy by comparison
of this method with another one, based on nebular SN~Ia spectra. 

Taking $L_{\rm max} \approx 4.3\times 10^{42} $ erg s$^{-1}$ 
for SN~2005hk (see Figure~\ref{fig:uvoir}) and the relation 
$L_{\rm max}= 2\times 10^{43} M_{\rm Ni}/M_{\sun}$, we obtain the crude estimate of
$M_{\rm Ni} \approx 0.22M_{\sun}$ cited in \S~\ref{sec:res_uvoir}.  We do not add 
the 10\% correction here since the procedure of calculating UVOIR luminosity 
in the present work should produce numbers closer to the true $L_{\rm bol}$.  
This estimate of $M_{\rm Ni}$ indicates that the closest model from 
the set of synthetic light curves for deflagration models computed by 
Blinnikov et al. (2006) is the one with the lowest $^{56}$Ni~mass, namely 
\emph{1\_3\_3}.  Details of the construction and flame simulation for this 
model are described by \citet{roepke2006}\footnote{Note that the model 
parameters for \emph{1\_3\_3} were somewhat unorthodox: the metallicity
was three times solar, the carbon mass fraction was 62\%, and the central
density at ignition was lower than usual.}.
The model \emph{1\_3\_3} has $M(^{56}\textrm{Ni}) =0.24 M_{\sun}$ and rather 
low asymptotic kinetic energy $E_{\rm kin}=0.365$ 
foe\footnote{1 foe $ = 10^{51}$ erg.}.
The details of the radiation hydrodynamics 
code {\sc stella} used for computation of theoretical
light curves and relevant references are given by \citet{Bli_etal06}.

Figure~\ref{fig:uvoir} presents the synthetic bolometric luminosity 
$L_{\rm bol}$ and Figure~\ref{fig:modellc} illustrates the 
$u'BVg'r'i'YJHK$ light curves computed from the multi-group fluxes of the 
model \emph{1\_3\_3}.  Zero time here is the moment of explosion.  We give 
the results for this model without fine-tuning it to the observations and 
with all numerical noise visible as bumps and wiggles on the plots. 
However, note that the synthetic fluxes here are convolved with the set
of filter functions actually used for observations in this work.
The only free parameter is the explosion epoch, which we have assumed to be
15~days before $B$ maximum, consistent with the discussion in 
\S~\ref{sec:res_optlc}.  Figure~\ref{fig:uvoir} shows that this epoch provides 
an excellent fit to the rising part of $L_{\rm bol}$, as well as the first 
$\sim$20~days of the post-maximum decline.  At later epochs, the model 
declines somewhat more slowly than the observations.  A better fit could be 
produced, probably, by a model with a little less $^{56}$Ni.

Figure~\ref{fig:modellc} illustrates that, without any tuning of the 
underlying hydrodynamic model, the fluxes in the $BVg'$ filters resemble the
observations reasonably well for the first 40 days, except for the fact 
that the peaks of the light curves occur $\sim$2 days earlier than those of
the models.  Without a detailed new hydrodynamic and nucleosynthetic model, 
we may only speculate that a faster evolution in $u'BVg'r'$ and on the tail
of the bolometric light curve could be obtained by reducing the opacity 
in the iron-peak elements.  Since diffusion is determined not only by opacity,
but also by density, this might be possible if the kinetic energy of the 
explosion were the same, or a bit higher, than in the plotted model.  If the 
reduction of $^{56}$Ni were due to lower consumption of C/O in general, 
leading to a reduction of all products of burning (Fe, Si, S, etc.), then the 
kinetic energy would also be reduced, the velocity would be lower, and at each 
moment of time the density would be higher in each Lagrangean layer.  This 
would likely lead to a slower, rather than a faster, evolution of the light 
curves.  However, if the reduction of $^{56}$Ni is accomplished by a burning 
mode which produces, say, more Si at the expense of iron-peak elements, then 
the kinetic energy would be almost the same (since the energy released per unit
mass would be almost the same) or even higher (if the amount of Si produced 
is greater than the corresponding reduction of iron-peak elements because more 
total C/O is consumed).  This would, in turn, lead to a lower opacity at 
the same (or even lower) density -- which is exactly what is required to 
produce faster diffusion and faster light curves.  To our knowledge, the 
feasibility of such a model does not contradict any fundamental properties of 
deflagrations in Chandrasekhar mass progenitors.

The deviations in $BVg'$ become appreciable two months after the 
explosion when the observed light curves become flatter.  There may be two 
reasons for this.  At this epoch the ejecta become almost transparent in 
visible light and the local thermodynamic equilibrium (LTE) 
assumption employed by {\sc stella} becomes less 
and less reliable.  Another cause of the flattening of the light curves may 
be stronger concentration of density in the ejecta (and of initial $^{56}$Ni).  
Then the flattening may be obtained even in the LTE approximation as we know
from the experience of radiation modeling of hundreds of ``toy'' SN~Ia models.
See the discussion of those models in the recent paper by  \citet{Woosley07}.

The deviations in $i'$ and NIR bands are generally more appreciable.  Only 
the general trends -- the flattening of the light curves after the maxima and
the absence of secondary maxima -- are reproduced.  As discussed by 
\citet{Woosley07}, when the results 
produced by {\sc stella} for the ``toy'' SN~Ia models are compared with the quite 
independent radiative transfer code {\sc sedona}, the behavior in the red and 
near-infrared is very sensitive to the lists of lines used for computation of 
the opacity.  While {\sc stella} employs $\sim 2\times 105$ lines which are 
mostly in the ultraviolet, it is necessary to use millions of weak lines in 
cool ejecta.  The work of inclusion of extended line lists into the 
{\sc stella} algorithm is in progress.

A plot of the expansion velocity of the ``Rosseland'' photosphere of model
\emph{1\_3\_3} is compared with measurements of the \ion{Si}{2} $\lambda$6355 
line in Figure \ref{fig:02cx_05hk_vels}.  If, alternatively, we were to define 
the photosphere as the level with an optical depth of 2/3 in the $BV$ bands, a 
somewhat flatter evolution of the expansion velocities would be predicted 
\citep[see][]{Bli_etal06}.  However, over the range of epochs that the 
$\lambda$6355 line was visible in the spectrum of SN~2005hk, both definitions 
give similar results.  We note that within a fairly short time ($\leq$2 weeks)
after explosion, the rapidly expanding ejecta of a SN~Ia become optically thin 
in the continuum, making the concept of a photosphere rather 
ambiguous.  Hence, the interpretation of such comparisons between models 
and observations is not completely straightforward.  Nevertheless, the 
agreement between model \emph{1\_3\_3} and our SYNOW estimate of the evolution 
of the photospheric velocity of SN~2005hk is encouraging, and 
suggests that a model with somewhat lower photospheric velocity than in
\emph{1\_3\_3} (e.g., again due to a stronger concentration of initial 
$^{56}$Ni) would provide a better fit to SN~2005hk.

Finally, we comment on the peculiar high-ionization SN~1991T-like pre-maximum 
spectra of this event.  Although we have not yet tried to model the spectra,
it is quite natural to expect the early high ionization in the deflagration 
models
since $^{56}$Ni lumps are predicted to float in the outermost layers;
see, e.g., the latest hydrodynamic simulations of SN~Ia in \citet{roepke2007}.
The results of numerical experiments on 3D gamma-ray transport for MPA 
deflagration models show the $^{56}$Ni-blobs are ``visible'' in
gamma rays very early, hence one should expect the high ionization.

\section{DISCUSSION}
\label{sec:disc}

From the results of the preceding section, it appears plausible that the subset
of SN~2002cx-like SNe~Ia can be understood in terms of the pure deflagration of 
a Chandrasekhar-mass white dwarf.  Nevertheless, a potentially serious
discrepancy remains in the large amount of low-velocity unburned oxygen and 
carbon that is predicted by current 3D deflagration models. 
\citet{Koz_etal05} have shown that the late-time spectra of these models should
exhibit strong forbidden lines of \ion{O}{1} and \ion{C}{1}, yet these features
were not observed in spectra of SN~2002cx obtained 227 and 277~days after 
maximum \citep{Jha_etal06}.  \citet{Koz_etal05} pointed out that the discrepancy 
is reduced somewhat by improved initial conditions and higher resolution in the 
3D models.  

On the observational side, \citet{Jha_etal06} found evidence for 
an unexpectedly high mass and density at low velocity based on the presence of 
P-Cygni profiles of \ion{Fe}{2}, \ion{Ca}{2}, and \ion{Na}{1} lines in the 
late-time spectra of SN~2002cx, as well as the absence of forbidden lines 
other than [\ion{Ca}{2}] $\lambda\lambda$7291,7324 and a few possible 
[\ion{Fe}{2}] lines.  Such high densities would rule out the presence of 
observable [\ion{O}{1}] and [\ion{C}{1}] lines, even if there were significant 
amounts of unburned oxygen and carbon near the center.  Hence, the real
discrepancy may be that the observations show a higher density at late epochs 
than the models predict.  Until now, deflagration modeling has focused on 
matching the properties of normal luminous SNe~Ia.  It is essential that 
new models be built with the specific goal of exploring whether a pure 
deflagration can produce light curves and spectra that match both the 
early- and late-epoch properties of SN~2002cx-like events.
Likewise, we encourage observers to obtain further spectroscopy of these
interesting objects, especially at late epochs and in the NIR.

Several authors \citep[e.g., LFC,][]{Bra_etal04,Cho_etal06} have commented on
the similarity of the pre-maxima spectra of SN~2002cx-like and SN~1991T-like
supernovae. Indeed, both types of events appear to occur preferentially in 
spiral galaxies, and so it is natural to ask if these two subclasses of 
SNe~Ia share a similar origin.
\citet{Bra_etal06} classified both SN~2002cx and SN~1991T as ``shallow silicon"
SNe~Ia, but they also emphasized the differences between the two objects:
SN~1991T had a slow-declining light curve, whereas SN~2002cx was 
faster-declining and subluminous; the ejecta of SN~1991T had typical expansion
velocities, while the expansion velocities of SN~2002cx were nearly a
factor of two lower at maximum; SN~1991T possessed a normal late-time 
spectrum, whereas the spectrum of SN~2002cx at late epochs was dominated
by very low-velocity permitted lines of \ion{Fe}{2} and displayed a
continuum or pseudo-continuum.  To these we can add the fact that the
$I$-band light curve of SN~1991T displayed a typical secondary peak
\citep{For_etal93}, while that of SN~2002cx conspicuously did not.  
As previously mentioned, the presence of a secondary peak at red and NIR
wavelengths is indicative of the iron-peak elements
being concentrated in the central regions, whereas the lack of a secondary
peak implies significant mixing of the ejecta.  This last difference
seems crucial and, along with the luminosity difference, argues against 
SN~1991T-like events also being produced by deflagrations.  As 
\citet{Bra_etal06} have speculated, it may be that these two subclasses of 
SNe~Ia have little in common except a high initial temperature.

The Lick Observatory Supernova Search has discovered (or recovered) 
four of the five 
known SN~2002cx-like SNe (2003gq, 2005P, 2005cc, and 2005hk), and we can use
this survey to estimate the frequency of this subclass among the general
SN~Ia population.  According to \citet{Li_etal01}, LOSS discovers 
essentially all of the SNe~Ia, including the subluminous SN~1991bg-like objects, 
that appear in galaxies in a distance-limited sample where normal SNe~Ia reach 
an apparent $R$-band magnitude at maximum of $\sim$16 or brighter.  If we take 
the absolute magnitude of a normal SN~Ia in $R$ to be $-19.4$ mag, this corresponds 
to a distance of $\sim$120~Mpc.  The average of the $R$-band absolute magnitudes 
of SN~2002cx and SN~2005hk was $-17.9$, which is very similar to that of 
typical SN~1991bg-like events (see Table~\ref{tab:absmags}).  Thus, LOSS 
should also discover most of the SN~2002cx-like events within $\sim$120~Mpc.
Since 2002, LOSS has discovered $\sim$90 SNe~Ia out to this
distance, of which 4 were confirmed SN~2002cx-like 
events.  These objects therefore appear to account for 
$\sim$5\% of all SNe~Ia in the local Universe, although there are at least
three caveats associated with this estimate.  First, there are certain
biases in the LOSS sample galaxies -- e.g., there aren't many irregular/dwarf
galaxies -- so our statistics may be biased depending on the frequency of
SN~2002cx-like events in such galaxies.  Second, we cannot say with
complete certainty that all of the SN~2002cx-like SNe in the LOSS discoveries
since 2002 have been identified, so this estimate is a lower limit.
Finally, the frequency of SN~2002cx-like SNe would be larger if there exists a 
significant population with absolute magnitudes considerably fainter than 
$M(R) = -17.9$ mag.

The high degree of similarity of the spectra and light curves of SN~2002cx 
and SN~2005hk is striking, and leads us to wonder how homogeneous this
subclass of SNe~Ia may actually be.  Is there something special about the
conditions required for a C/O white dwarf to explode as a pure deflagration
that leads to such homogeneity, or is the twin-like resemblance of these two 
SNe merely a coincidence?  Due to the rarity of these events, answering this 
question requires detailed spectroscopic and photometric observations of 
large samples of nearby SNe.  Such data do not yet exist, but will eventually
become available due to the combined efforts of groups such as the CSP, 
SDSS~II, LOSS, the CfA Supernova 
Group\footnote{http://www.cfa.harvard.edu/oir/Research/supernova/},
and the Nearby Supernova Factory \citep{Ald_etal02}.  These large surveys may
also tell us more about the possible progenitors of this peculiar
subclass of SNe~Ia.  As mentioned, all five of the SN~2002cx-like 
objects discovered to date occurred in spiral galaxies showing clear evidence
of ongoing star formation.  Do SN~2002cx-like SNe also occur in elliptical or S0
galaxies?  If not, this may indicate that the conditions for pure 
deflagrations are produced only in a younger or intermediate-aged stellar 
population.  At present, however, such ideas are pure speculation.

Finally, it is interesting to ask whether SN~2002cx-like events could be 
mistakenly
included in high-redshift samples of normal SNe~Ia.  Both SN~2002cx and 
SN~2005hk were more than a magnitude fainter in absolute magnitude in the 
$B$ and $V$ bands than normal SNe~Ia.  The Supernova Legacy Survey, which 
discovers SNe~Ia in the redshift range $z \approx 0.3$--1.0 and is 
representative of current ground-based high-redshift SN surveys, has detected 
SNe~Ia down to $M(B) \approx -19.0$ mag \citep{Ast_etal06}.  Hence, if SN~2002cx
and SN~2005hk are representative of the subclass of SN~2002cx-like events as
a whole, it is unlikely that any of these objects would be discovered in 
high-redshift surveys.  In any case, if an adequate spectrum is obtained 
near maximum, the low expansion velocities that typify this peculiar 
subclass of SNe~Ia should serve to identify them.

\acknowledgments 

We are grateful for the assistance of the staffs at the many 
observatories (Lick, Keck, LCO, MDM, APO, ESO) where data for 
this paper were obtained.
We would like to thank David Branch for allowing us to use SYNOW, and 
Jerod Parrent for providing us the code.  This paper is based upon 
CSP observations supported by the National Science Foundation under Grant 
No. 0306969.  
S.~Blinnikov is supported in Russia partly by grants RFBR 05-02-17480, 
04-02-16793, RFNS 540.2006.2, 10181.2006.2, and by grant 
IB7320-110996/1 of the Swiss National Science Foundation.
M. Hamuy acknowledges support from the Centro de Astrof\'isica
FONDAP 15010003 and to Proyecto Fondecyt 1060808.
The authors made use of the 
NASA/IPAC Extragalactic Database (NED), which
is operated by the Jet Propulsion Laboratory, California Institute of
Technology, under contract with the National Aeronautics and Space 
Administration.
A.V.F.'s group at U.C. Berkeley is supported by NSF grant AST-0607485
and the TABASGO Foundation.  KAIT
was made possible by generous donations from Sun Microsystems, Inc., the
Hewlett-Packard Company, AutoScope Corporation, Lick Observatory, the National
Science Foundation, the University of California, and the Sylvia \& Jim Katzman
Foundation.

Funding for the SDSS and SDSS~II has been provided by the Alfred P.  
Sloan Foundation, the Participating Institutions, the National  
Science Foundation, the U.S. Department of Energy, the National  
Aeronautics and Space Administration, the Japanese Monbukagakusho,  
the Max Planck Society, and the Higher Education Funding Council for  
England.
The SDSS is managed by the Astrophysical Research Consortium for the  
Participating Institutions. The Participating Institutions are the  
American Museum of Natural History, Astrophysical Institute Potsdam,  
University of Basel, Cambridge University, Case Western Reserve  
University, University of Chicago, Drexel University, Fermilab, the  
Institute for Advanced Study, the Japan Participation Group, Johns  
Hopkins University, the Joint Institute for Nuclear Astrophysics, the  
Kavli Institute for Particle Astrophysics and Cosmology, the Korean  
Scientist Group, the Chinese Academy of Sciences (LAMOST), Los Alamos  
National Laboratory, the Max-Planck-Institute for Astronomy (MPIA),  
the Max-Planck-Institute for Astrophysics (MPA), New Mexico State  
University, Ohio State University, University of Pittsburgh,  
University of Portsmouth, Princeton University, the United States  
Naval Observatory, and the University of Washington.

\appendix

\section{Re-reduction of $BVRI$ Photometry of SN~2002cx}

The $BVRI$ photometry of SN~2002cx published by LFC suffered 
from contamination by the underlying host-galaxy light due to the lack of
appropriate template images for subtracting the galaxy.  In this Appendix,
we present a re-reduction of these data using host-galaxy template images 
obtained with both telescopes after the supernova had completely disappeared.

The LFC photometry of SN~2002cx was obtained with KAIT and 
1.0~m Nickel telescope at Lick Observatory.  Template $BVRI$ images of the 
host galaxy, CGCG~044-035, were obtained with both telescopes and the
same cameras and filters in March 2005.
Galaxy subtractions were performed on all images, and magnitudes for the
SN were measured from PSF-fitting photometry carried out differentially with 
respect to the comparison stars identified in Figure~1 of LFC.  The $BVRI$ 
magnitudes of the comparison stars were established from observations of 
\citet{lan92} standards carried out with KAIT on 13 photometric nights.
Uncertainties for the SN magnitudes were estimated from measurements made on 20
artificial stars having the same magnitude as the SN that were randomly
inserted in the images and re-extracted \citep{gan07}.  The final uncertainty was
taken to be the magnitude scatter of the artificial stars added in quadrature
with the uncertainty from the PSF photometry.

Table~\ref{tab:bvri02cx} lists the final revised $BVRI$ magnitudes for
SN~2002cx.

\clearpage
\figcaption[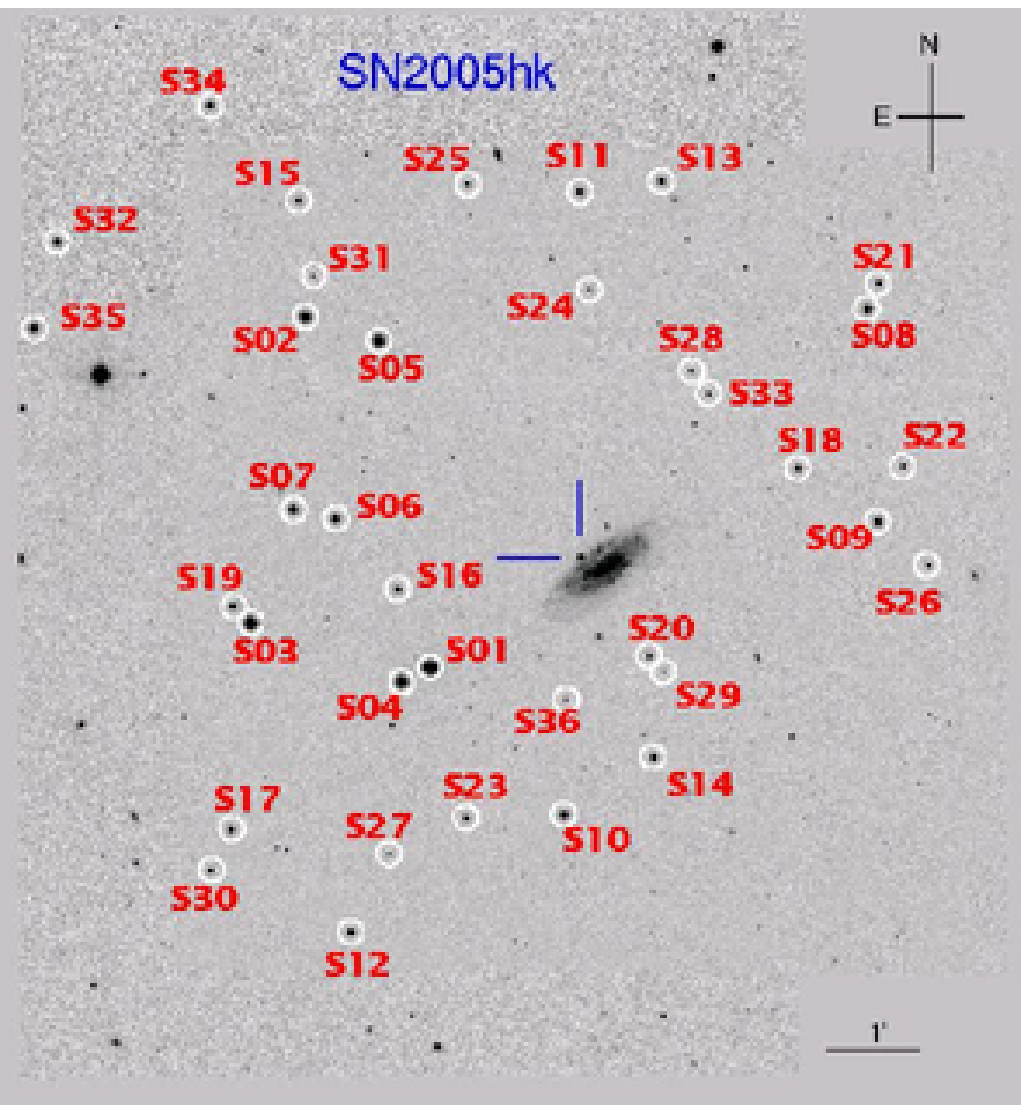]{The field of SN~2005hk observed with the Swope 1~m 
  telescope at
  LCO.  This image was created by combining optical and infrared images so 
  that some of the stars used for the NIR photometry would be clearer. 
  North is up and east is to the left. The supernova is marked to the 
  NE of the host-galaxy nucleus. The comparison stars used to derive 
  differential photometry of the SN are labeled. The plate scale is shown 
  near the bottom right.\label{fig:fc}}   

\figcaption[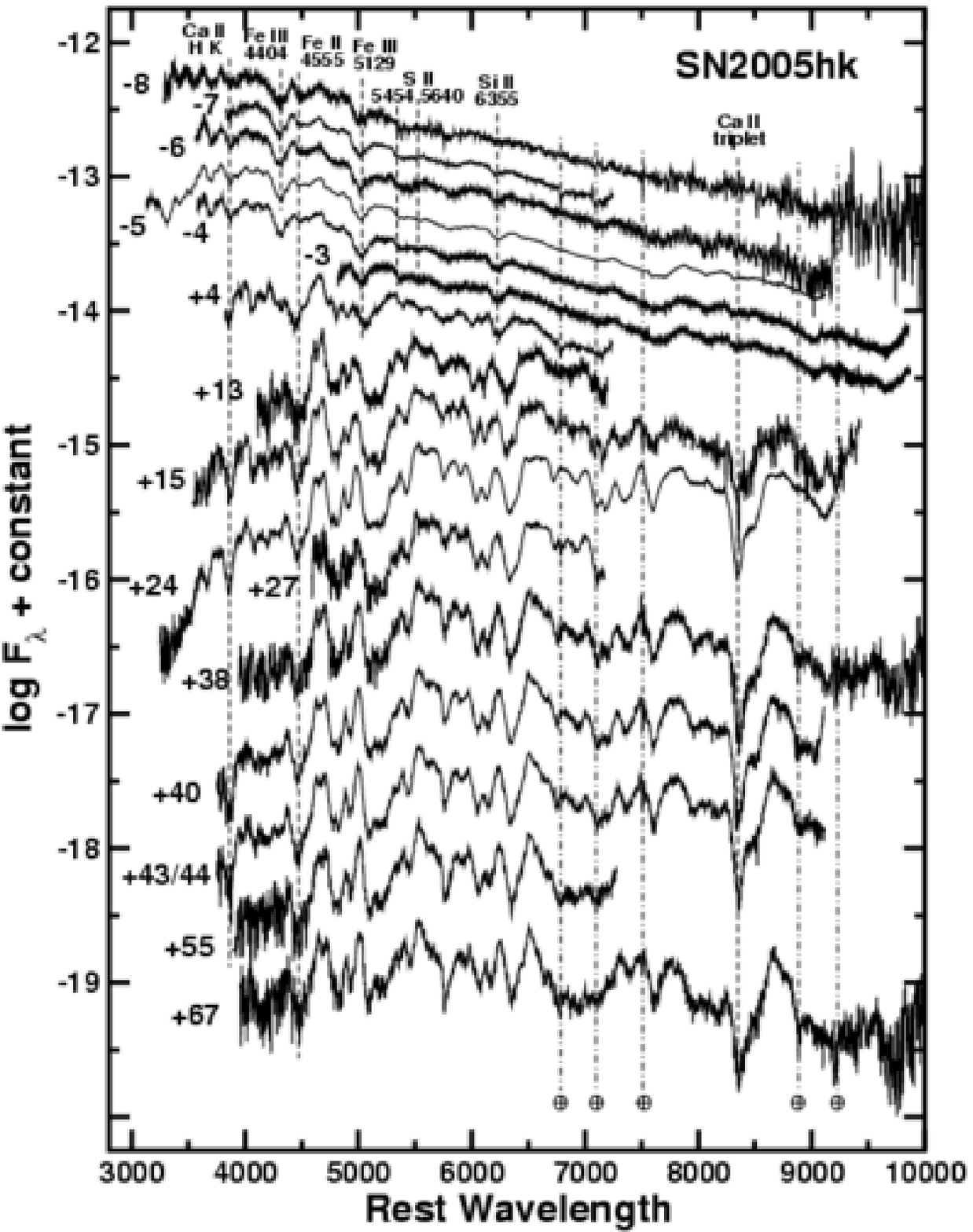]{Spectroscopic evolution of SN~2005hk. Each spectrum is
  plotted on a logarithmic flux-density scale, shifted by an arbitrary constant. 
  The wavelengths of the spectra were shifted to the SN rest frame using a 
  redshift of $z = 0.012993$ as given in NED. The labels to
  the left of each spectrum indicate the epoch in rest-frame days since
  the date of $B$ maximum (JD $=$ 2,453,685.1).  The spectra for days +43 and
  +44 have been averaged.  Identifications for selected spectral features 
  are indicated.  For most (but not all) of the spectra, the telluric 
  absorption features have been eliminated; the positions of the strongest
  of these are indicated by the $\Earth$ symbol.\label{fig:spmontage}}

\figcaption[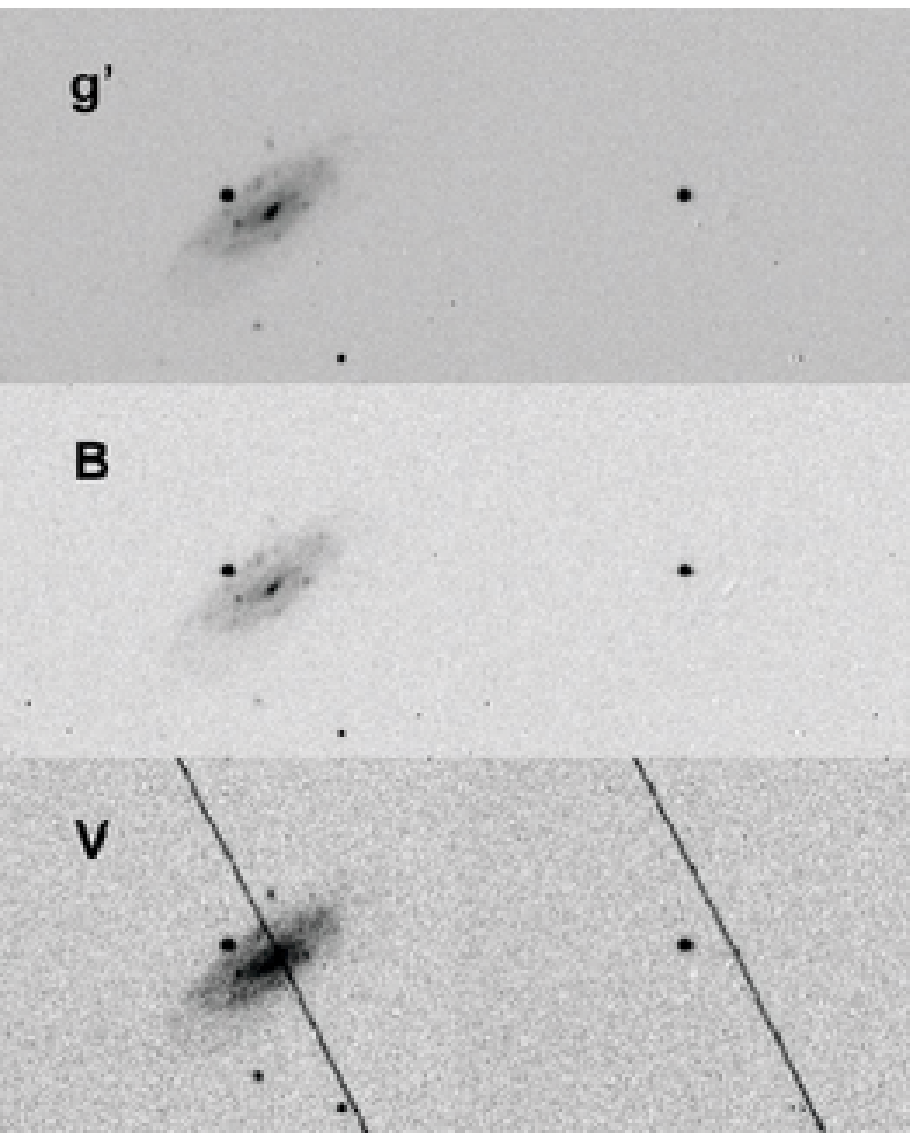]{Images of SN~2005hk in the $g'$, $B$, and $V$ bands 
  obtained on 
  2005~Nov.~08 by the CSP.  The images in the left half of the figure 
  are as observed; those on the right-hand side are shown after subtraction of 
  a template image taken by SDSS~II before the SN appeared.  Note the 
  satellite trail in the $V$ image, which serves as an indicator of the
  quality of the galaxy subtraction.  See text for
  more details.\label{fig:gBV}}  

\figcaption[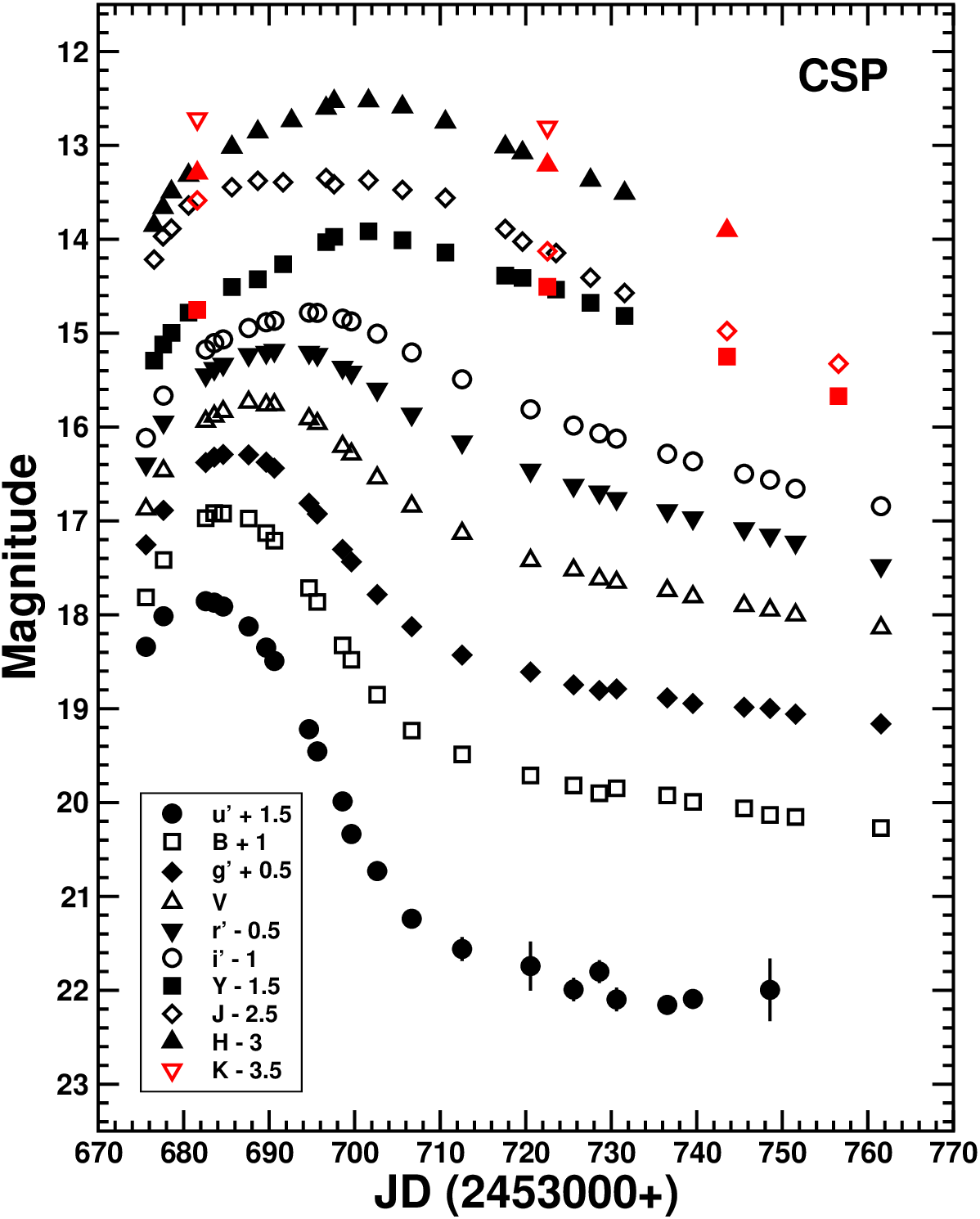]{Observed $u'g'r'i'BV$ and $YJHK$ light curves of 
  SN~2005hk obtained 
  by the CSP.  For clarity, the magnitudes in each band have been shifted by 
  an arbitrary constant which is given in the legend. The uncertainties in the
  measurements are smaller than the plot symbols except where indicated.
  $YJH$ photometry obtained with RetroCam is plotted with black symbols;
  $YJHK$ measurements made with WIRC are shown in red.
  \label{fig:csplc}}  

\figcaption[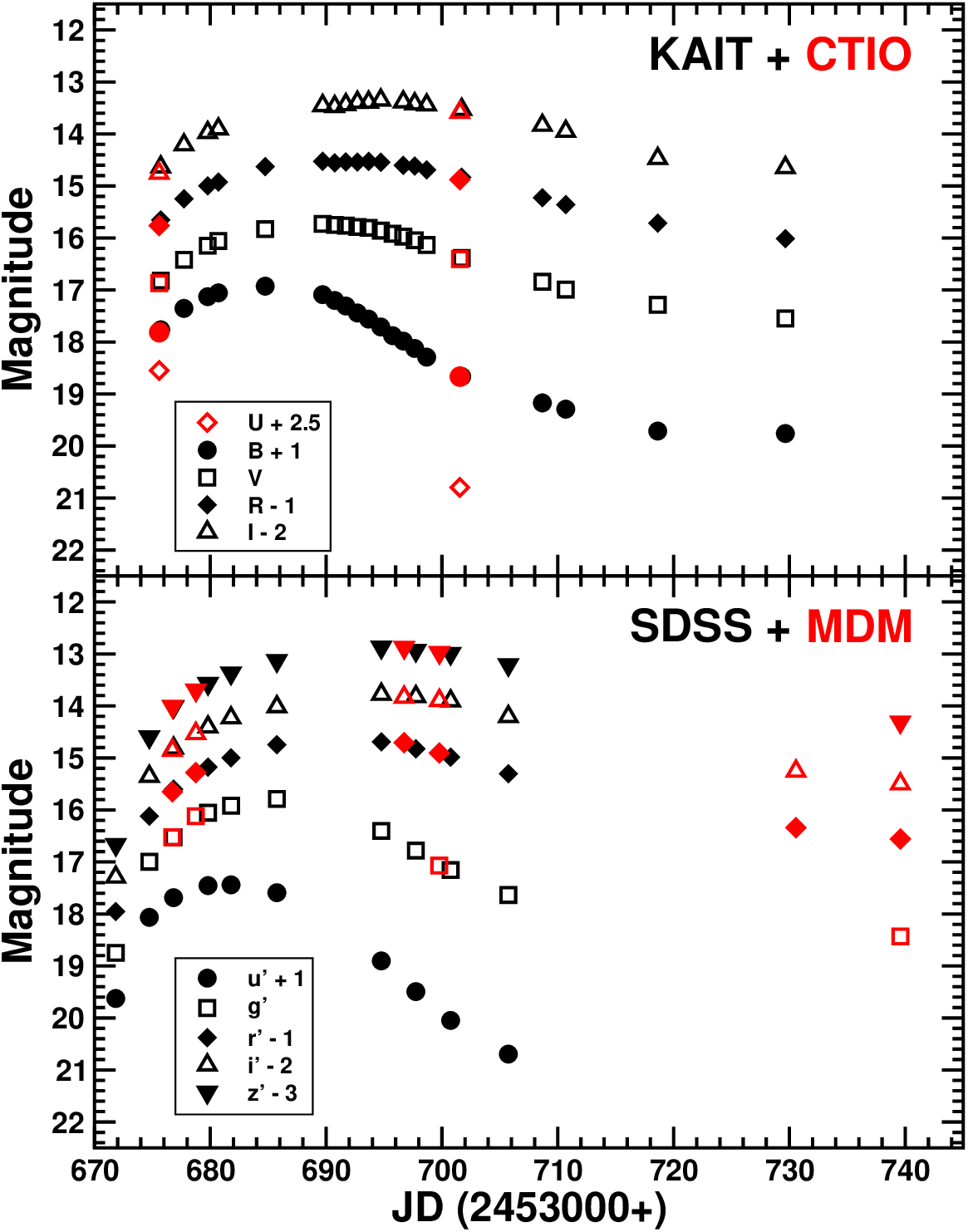]{(Top) Observed $BVRI$ light curves of SN~2005hk 
  obtained by KAIT and with the CTIO 0.9~m telescope.  
  The KAIT data are plotted with black symbols, and
  the CTIO with red.  Error bars are not shown as these are considerably
  smaller than the plot symbols.  No host-galaxy subtraction was performed 
  on the images before measurement.  (Bottom) Observed $ugriz$ light 
  curves of SN~2005hk obtained by the SDSS II Supernova Survey.  The SDSS
  and MDM photometry is derived from the scene modeling algorithm of
  \citet{Hol_etal07}.  Error bars are much smaller than the plot
  symbols.\label{fig:kaitsdsslc}} 

\figcaption[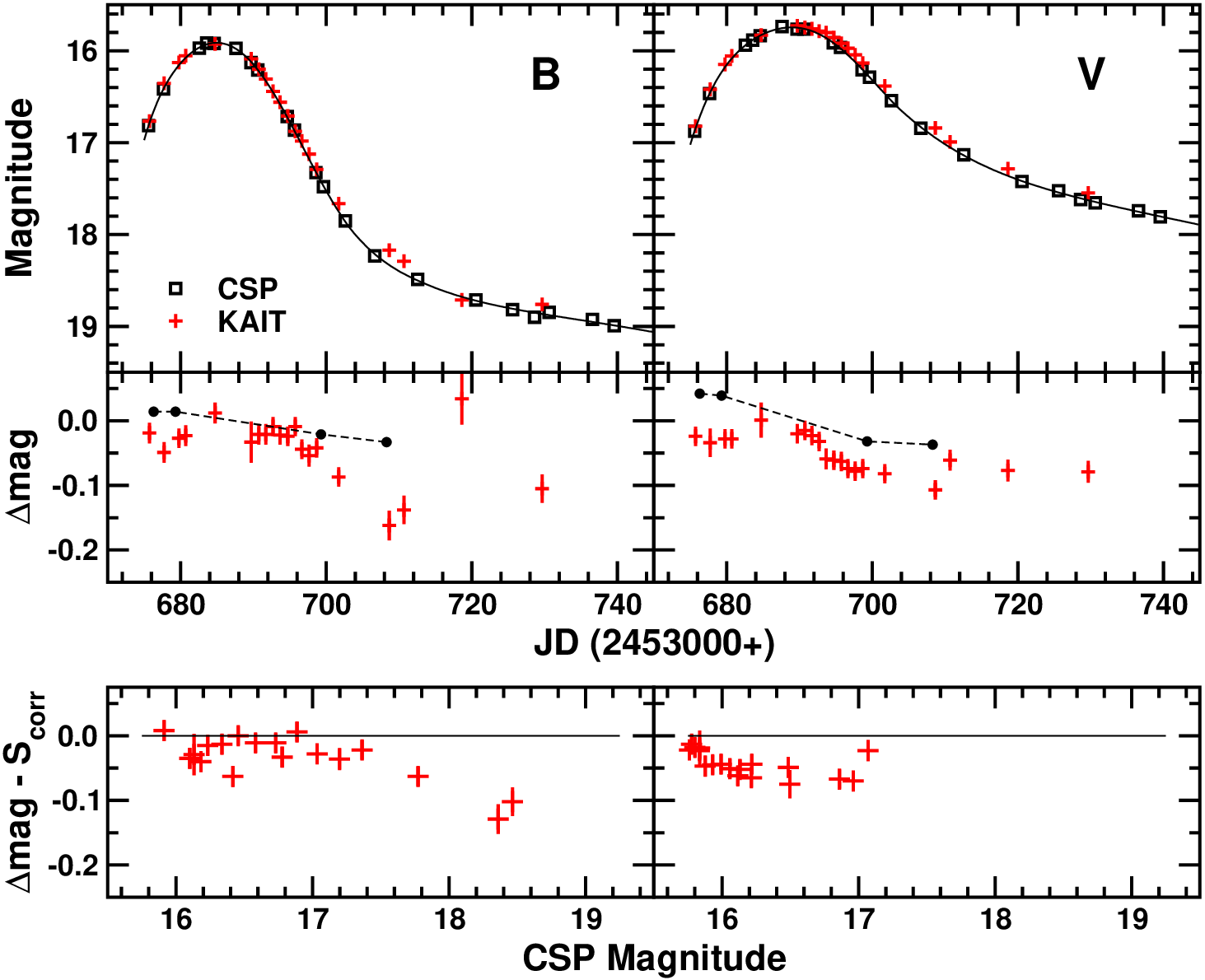]{Comparison of observed $BV$ light curves of SN~2005hk 
  obtained by the
  CSP and KAIT.  In the top panels, the square symbols correspond to the CSP
  measurements and the crosses to KAIT.  The solid line represents a fit to
  the CSP data.  In the middle panels, the difference between the KAIT 
  magnitudes and the smooth fit to the CSP data is shown.  Also indicated by
  the solid circles joined by dashed lines are the expected S-corrections
  calculated via synthetic photometry with the respective response functions
  of the CSP and KAIT passbands (see text for more details).  In the lower
  panels, the S-corrections have been subtracted from the magnitude differences
  between the KAIT and CSP measurements and then plotted vs. the CSP magnitude.
  \label{fig:cspkait}} 

\figcaption[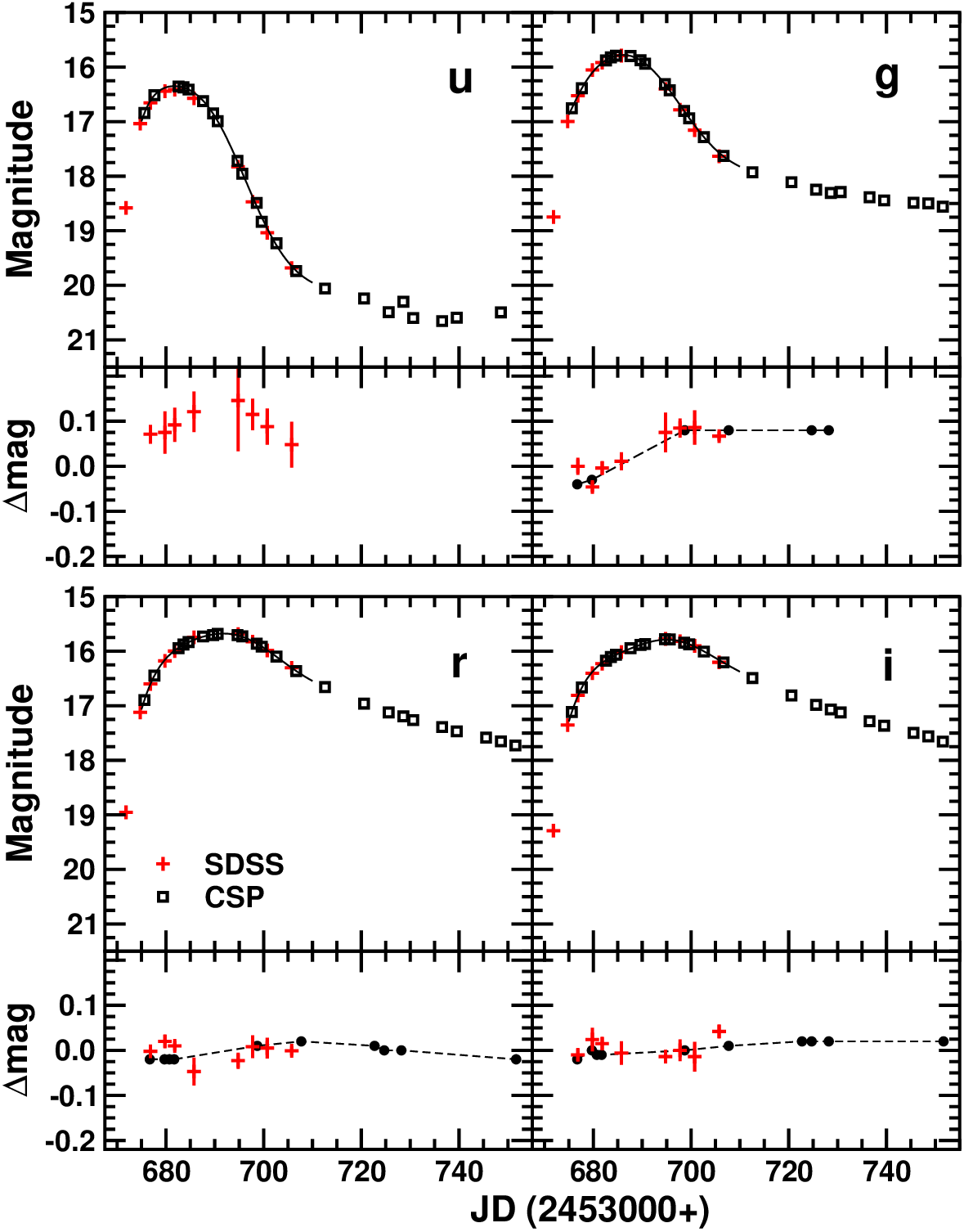]{Comparison of observed $ugri$ light curves of SN~2005hk 
  obtained
  by the CSP and SDSS~II.  In the top panels for each filter, the square 
  symbols correspond to the CSP measurements and the crosses to SDSS~II.  The 
  solid line represents a fit to the CSP data.  In the lower panels, the
  differences between the SDSS~II magnitudes and the smooth fit to the CSP data
  are shown.  These differences compare well with the S-corrections (indicated
  by the solid circles joined by dashed lines) calculated via synthetic 
  photometry from the respective response functions of the CSP and SDSS~II 
  passbands.\label{fig:cspsdss}} 

\figcaption[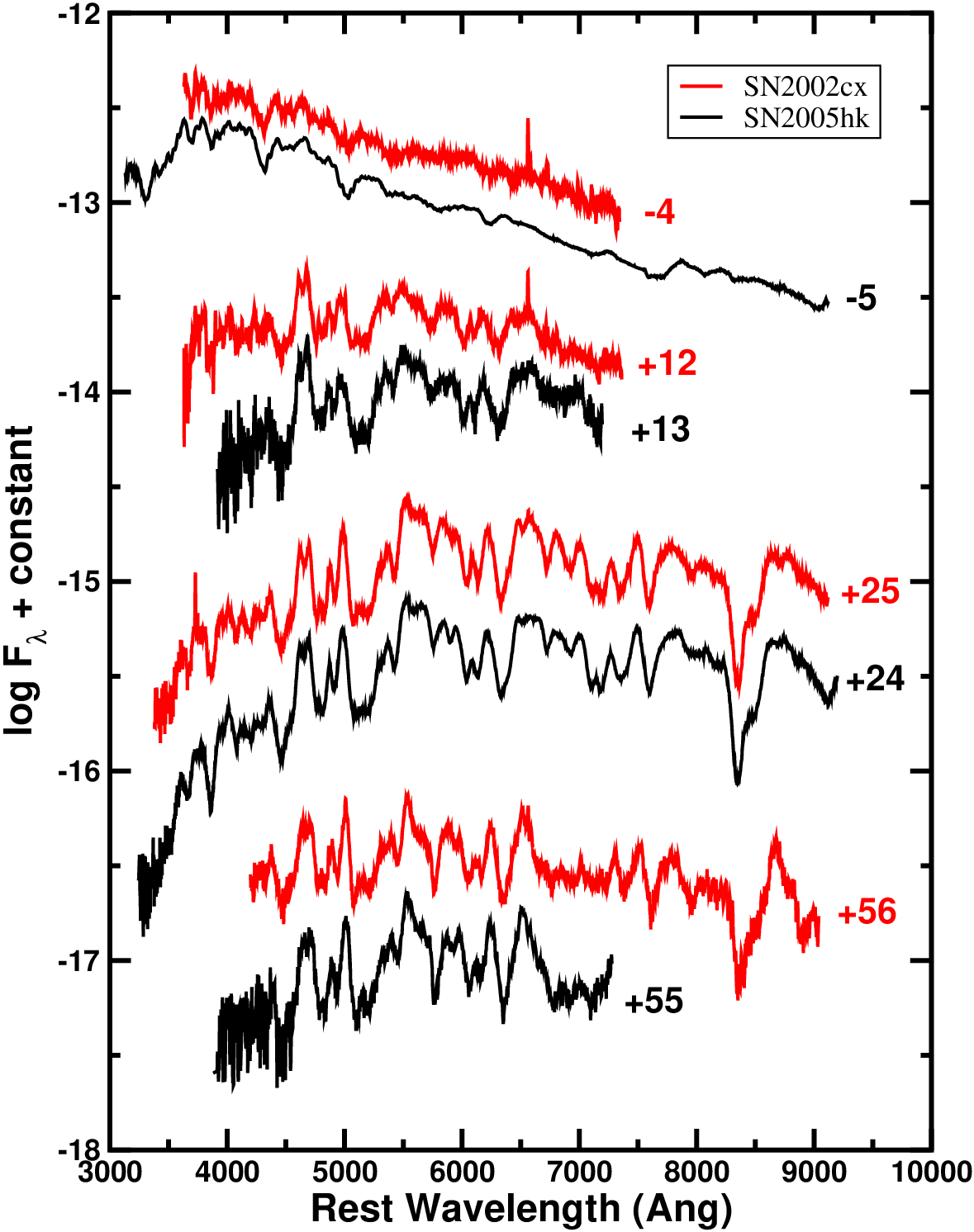]{Comparison of spectra of SN~2005hk at phases of $-5$, +13, 
  +24, and +55
  days with similar-epoch spectra of SN~2002cx from LFC.  The spectra are 
  plotted on a logarithmic flux scale and shifted by an arbitrary constant.  
  The wavelengths of the spectra were shifted to the SN rest frame using the
  heliocentric velocities of the host galaxies given in NED.
  \label{fig:02cx_05hk_spec}} 

\figcaption[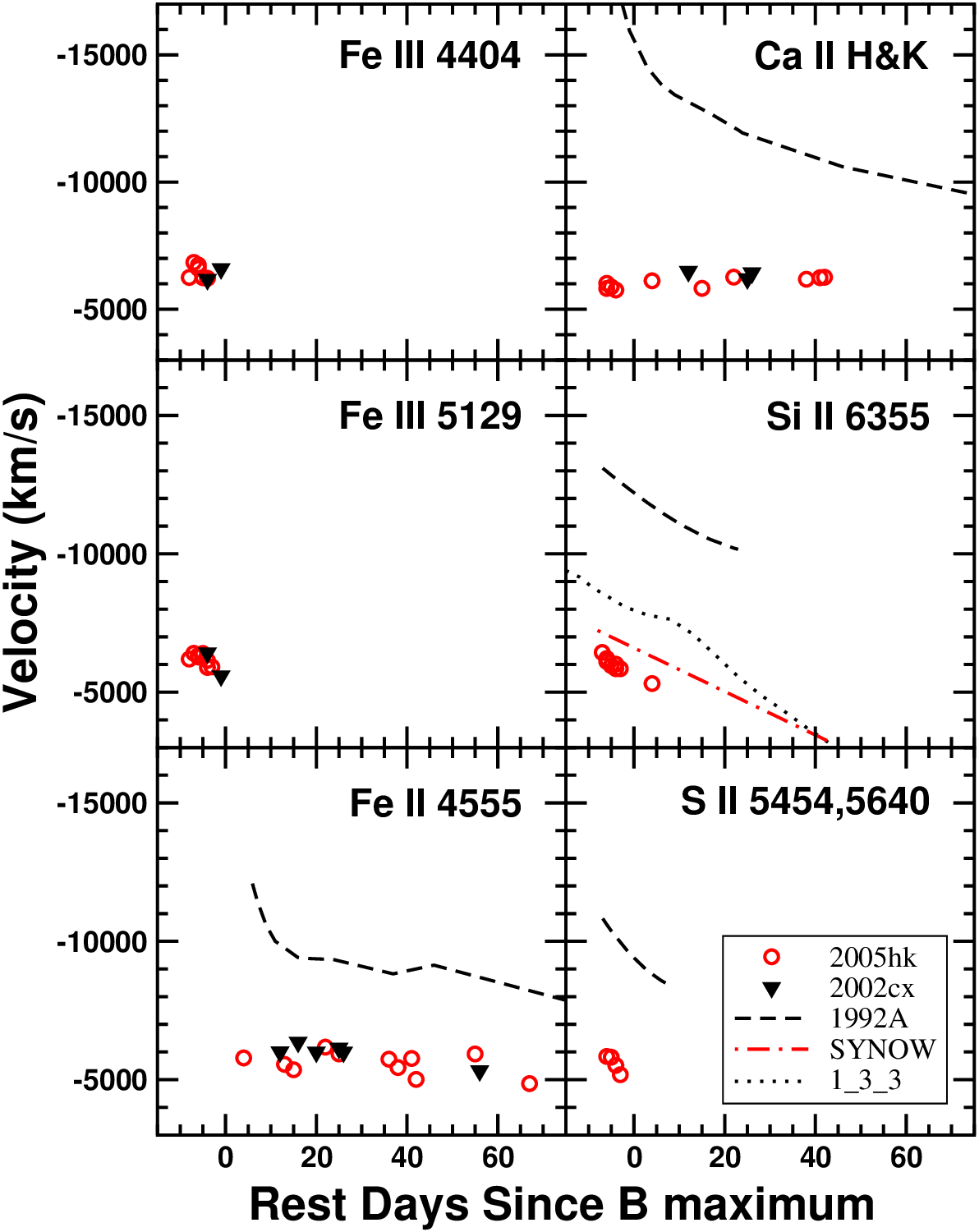]{Expansion velocities for SN~2005hk (circles) and 
  SN~2002cx (triangles)
  in selected lines of \ion{Fe}{2}, \ion{Fe}{3}, \ion{Ca}{2}, \ion{Si}{2}, and 
  \ion{S}{2}.  The measurements were estimated from the wavelength 
  of the minimum of each feature, correcting for the heliocentric velocities
  of the host galaxies.  For comparison, measurements in the \ion{Fe}{2}, 
  \ion{Ca}{2}, \ion{Si}{2}, and \ion{S}{2} lines are shown for SN~1992A, a 
  typical SN~Ia with a decline rate of $\Delta$m$_{15}$($B$)$ = 1.47$ mag.  The
  evolution of the photospheric velocity of SN~2005hk as estimated from SYNOW
  fits is compared with the \ion{Si}{2} $\lambda$6355 measurements.  Finally,
  in the same plot, the velocity of the ``Rosseland'' photosphere for the 3D 
  deflagration model \emph{1\_3\_3} is indicated.\label{fig:02cx_05hk_vels}} 

\figcaption[fig10.eps]{Comparison of KAIT $BVRI$ photometry of SN~2005hk 
  (red plus symbols) 
  and SN~2002cx (black circles).  The light curves have been normalized to the
  same peak magnitudes.\label{fig:02cx_05hk_phot}} 

\figcaption[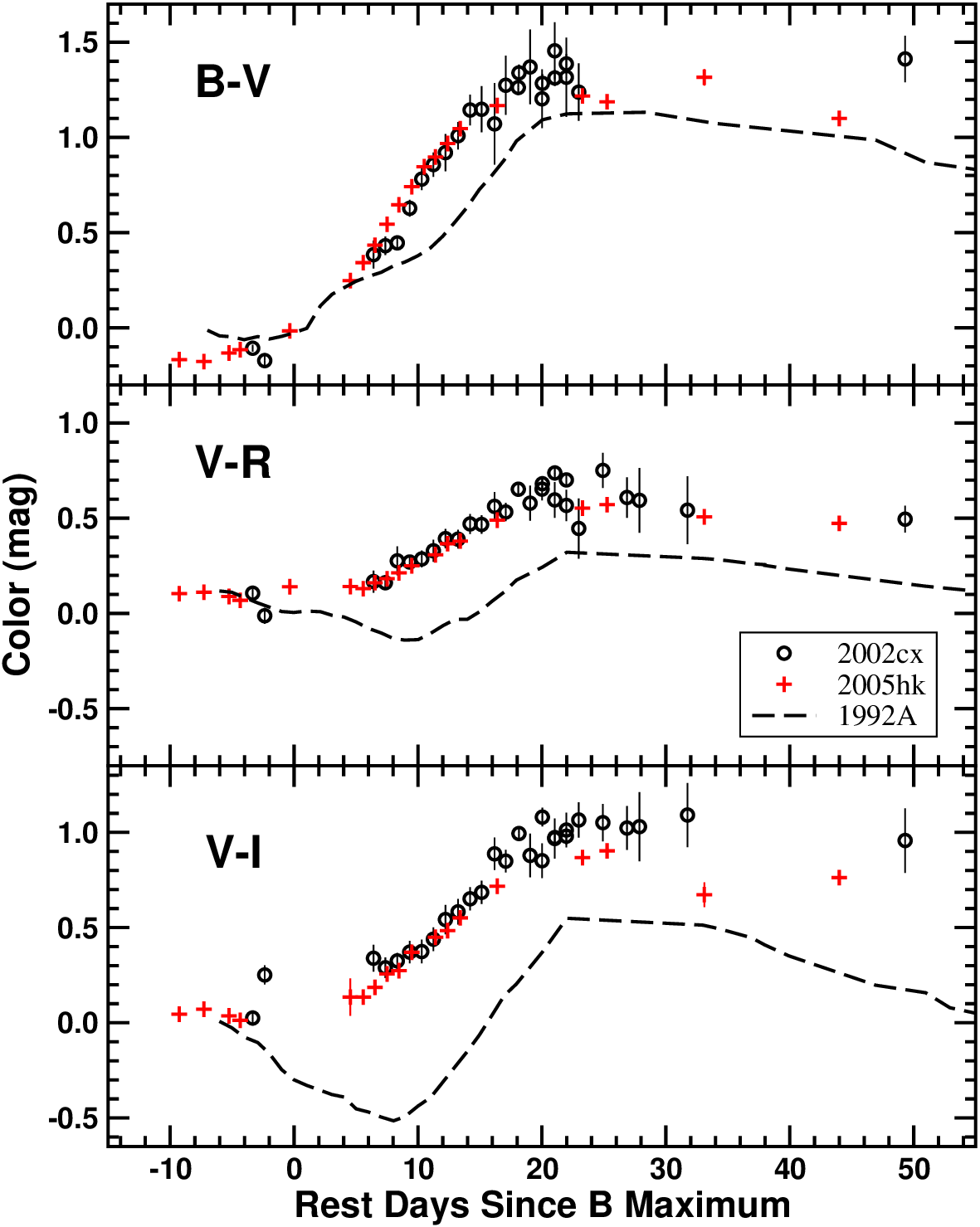]{Comparison of the $B-V$, $V-R$, and $V-I$ color 
  evolution of SN~2005hk 
  (red plus symbols) and SN~2002cx (black circles) as derived from KAIT 
  photometry.  The colors have been corrected for Galactic extinction assuming 
  the \citet{Sch_etal98} estimates of $E(B-V)_{\mathrm {Gal}} = 0.022$ mag for 
  SN~2002hk and 0.034 mag for SN~2002cx.  The colors of SN~2005hk have also been
  corrected for a host galaxy extinction of $E(B-V)_{\mathrm {Host}} = 0.09$ mag.
  The color evolution
  of SN~1992A, a typical unreddened SN~Ia with a decline rate of 
  $\Delta$m$_{15}$($B$)$ = 1.47$ mag, is indicated by the dashed line.
  \label{fig:02cx_05hk_colors}} 

\figcaption[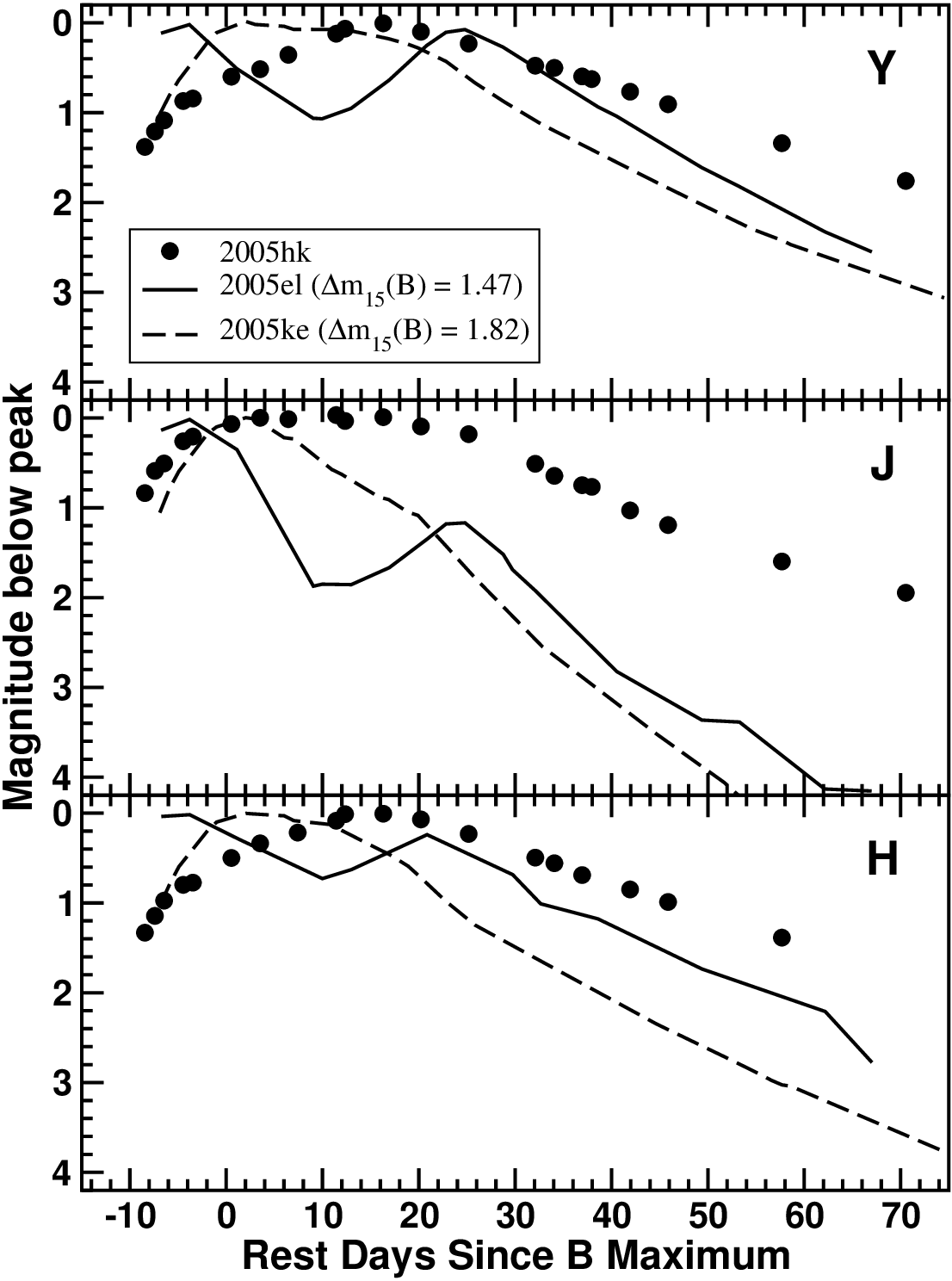]{Comparison of the CSP $JHK$ light curves of SN~2005hk 
  with CSP observations of SN~2005el ($\Delta$m$_{15}$($B$)$ = 1.47$ mag) and 
  SN~2005ke ($\Delta$m$_{15}$($B$)$ = 1.82$ mag).\label{fig:nir}} 

\figcaption[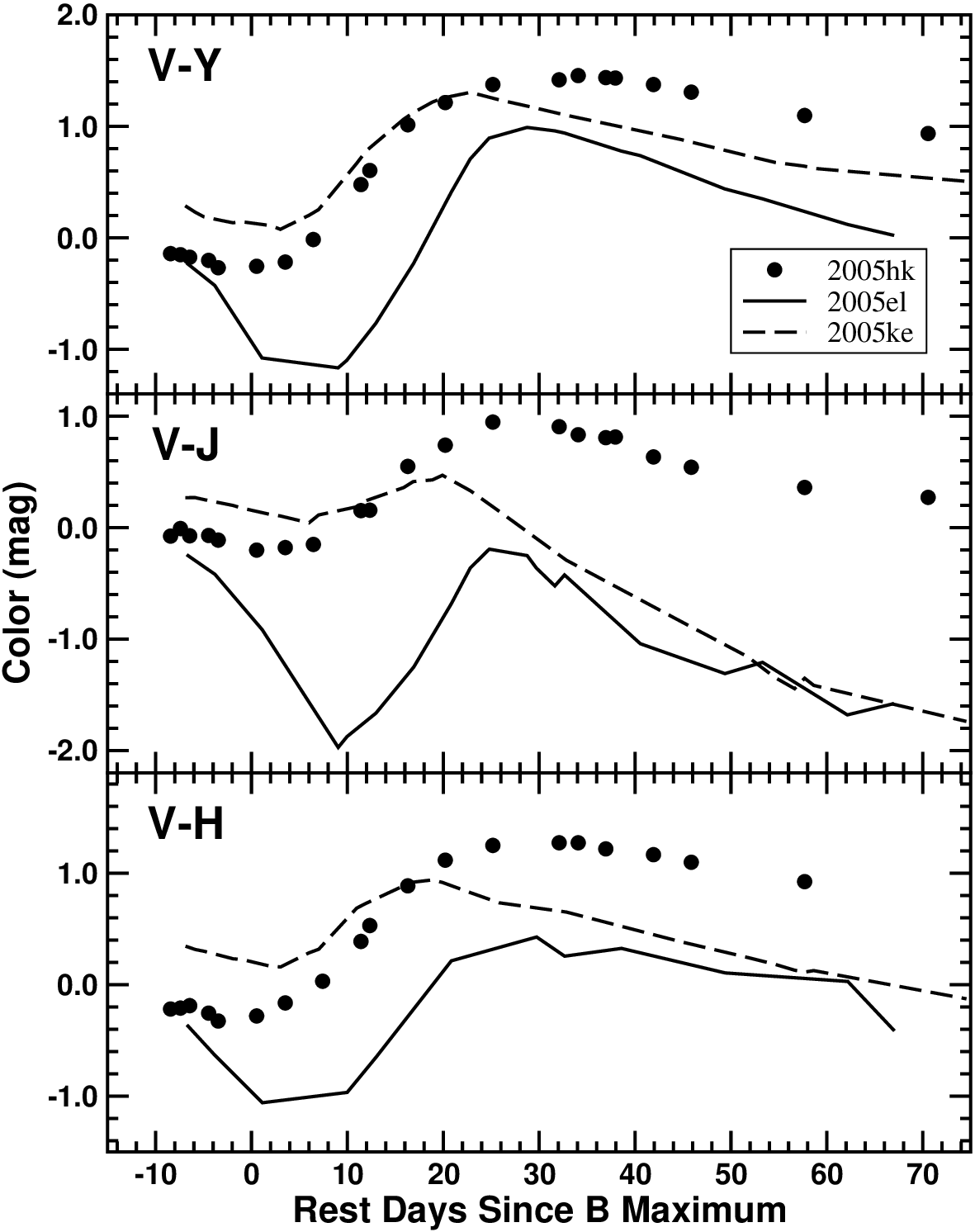]{Comparison of the $V-Y$, $V-J$, and $V-H$ color 
  evolution of SN~2005hk
  with CSP observations of SN~2005el ($\Delta$m$_{15}$($B$)$ = 1.47$ mag) and 
  SN~2005ke ($\Delta$m$_{15}$($B$)$ = 1.82$ mag).  The colors have been corrected
  for Galactic extinction using the estimates of \citet{Sch_etal98}.
  \label{fig:optir}} 

\figcaption[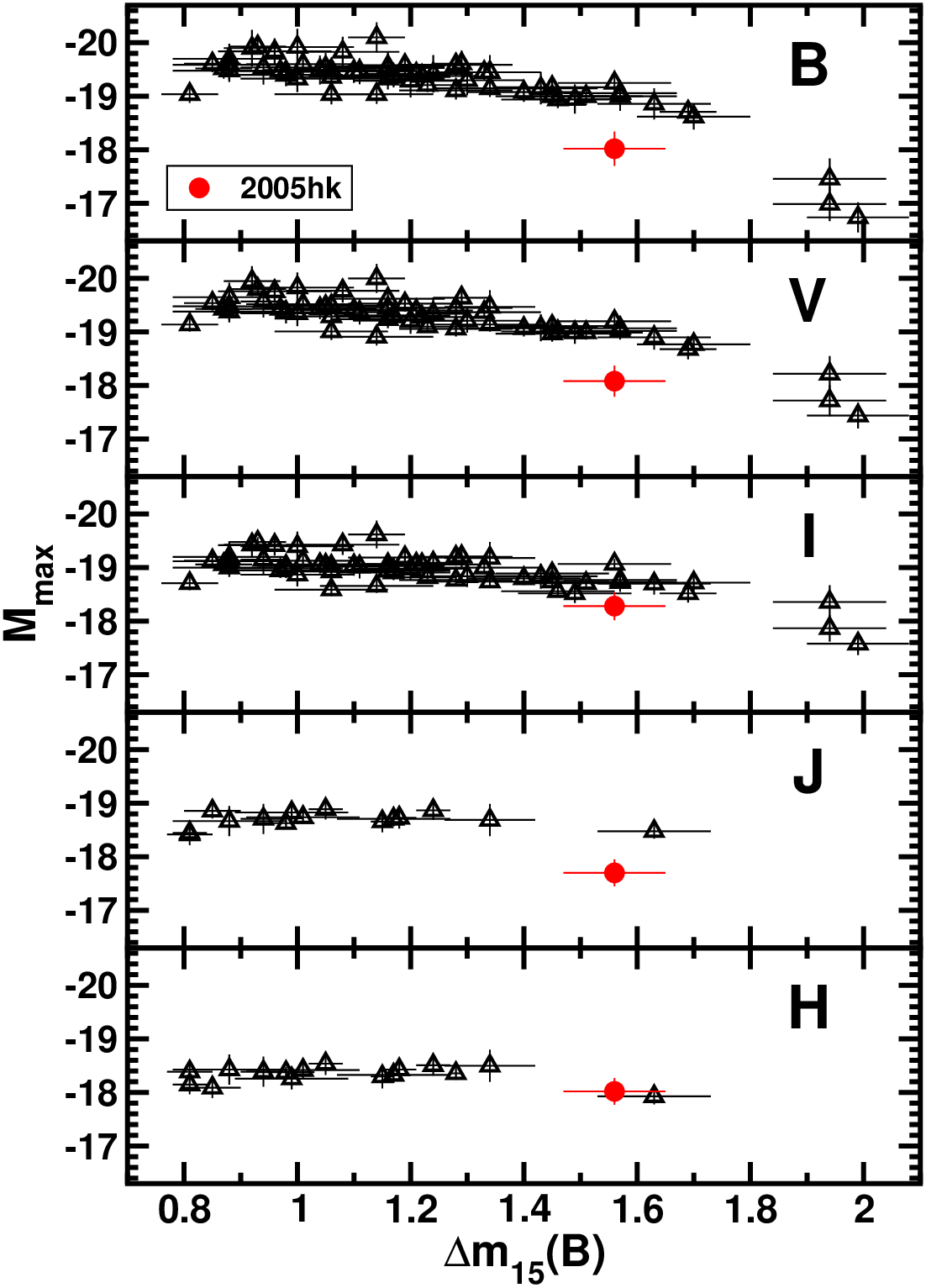]{The absolute magnitudes of SNe~Ia at maximum light 
  in the $BVIJH$
  bands plotted versus the decline-rate parameter $\Delta$m$_{15}$($B$).  
  The black triangles are SNe in the redshift range $0.01 < z < 0.1$ whose 
  distances were calculated from their host-galaxy radial velocities  
  in the cosmic microwave background frame assuming a Hubble constant of 
  $H_0 = 72$ km s$^{-1}$ Mpc$^{-1}$.  The red circle in each panel corresponds
  to SN~2005hk.\label{fig:pabsmags}} 

\figcaption[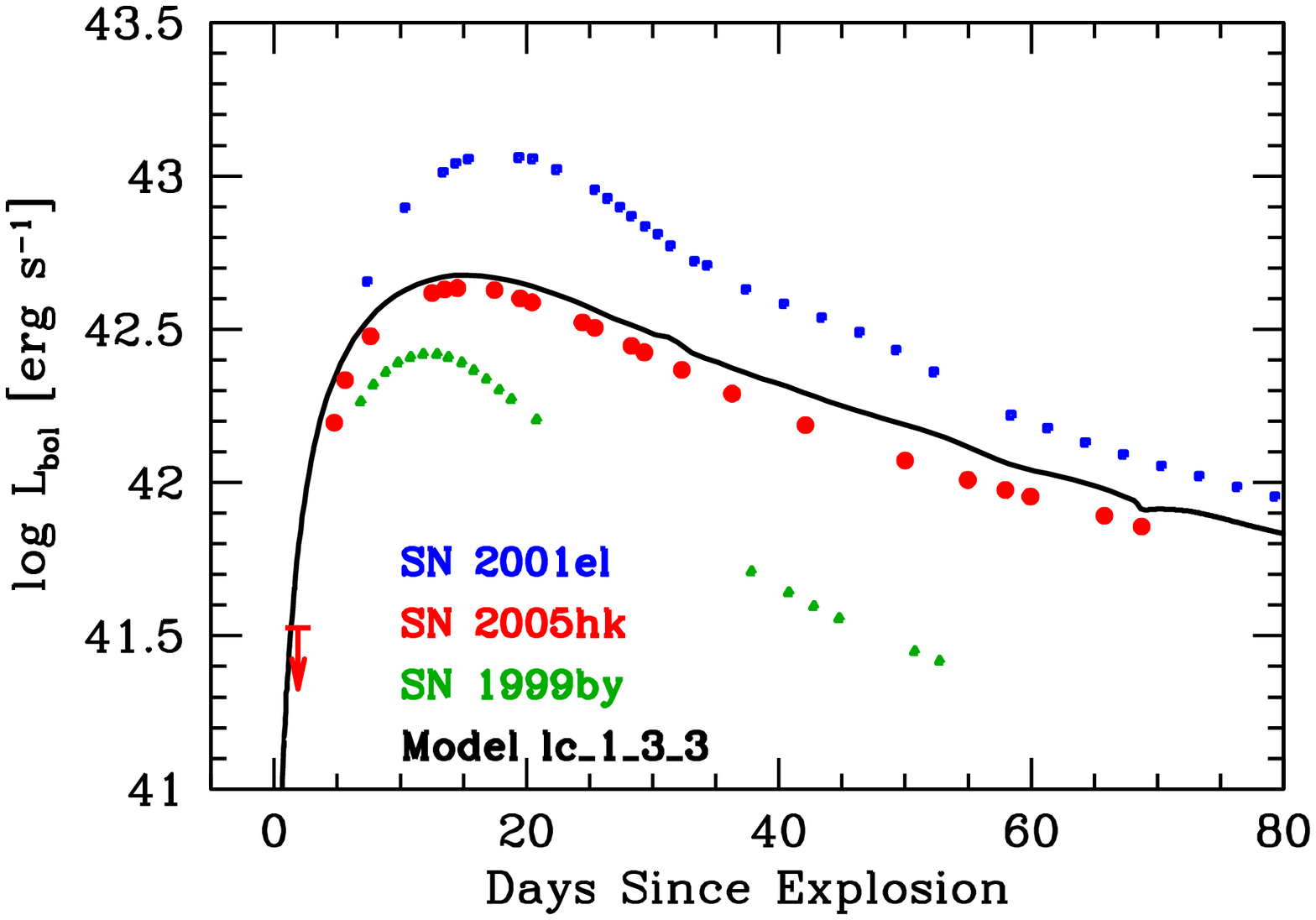]{UVOIR bolometric light curve of SN~2005hk plotted 
  with respect to the 
  epoch of explosion.  For reference, the UVOIR light curves of SN~2001el 
  ($\Delta$m$_{15}$($B$) = 1.13 mag) and SN~1999by ($\Delta$m$_{15}$($B$) = 
  1.90 mag) are also plotted as blue squares and green triangles, respectively.  
  For SN~2005hk, the explosion is assumed to have occurred 15~days before 
  $B$ maximum, whereas for SN~2001el and SN~1999by this number is take to
  be 18.1 and 11.0~days before $B$ maximum, respectively.  The data for 
  SN~2005hk are compared with the UVOIR light
  curve for the 3D deflagration model \emph{1\_3\_3}.  See text for further 
  details.\label{fig:uvoir}} 

\figcaption[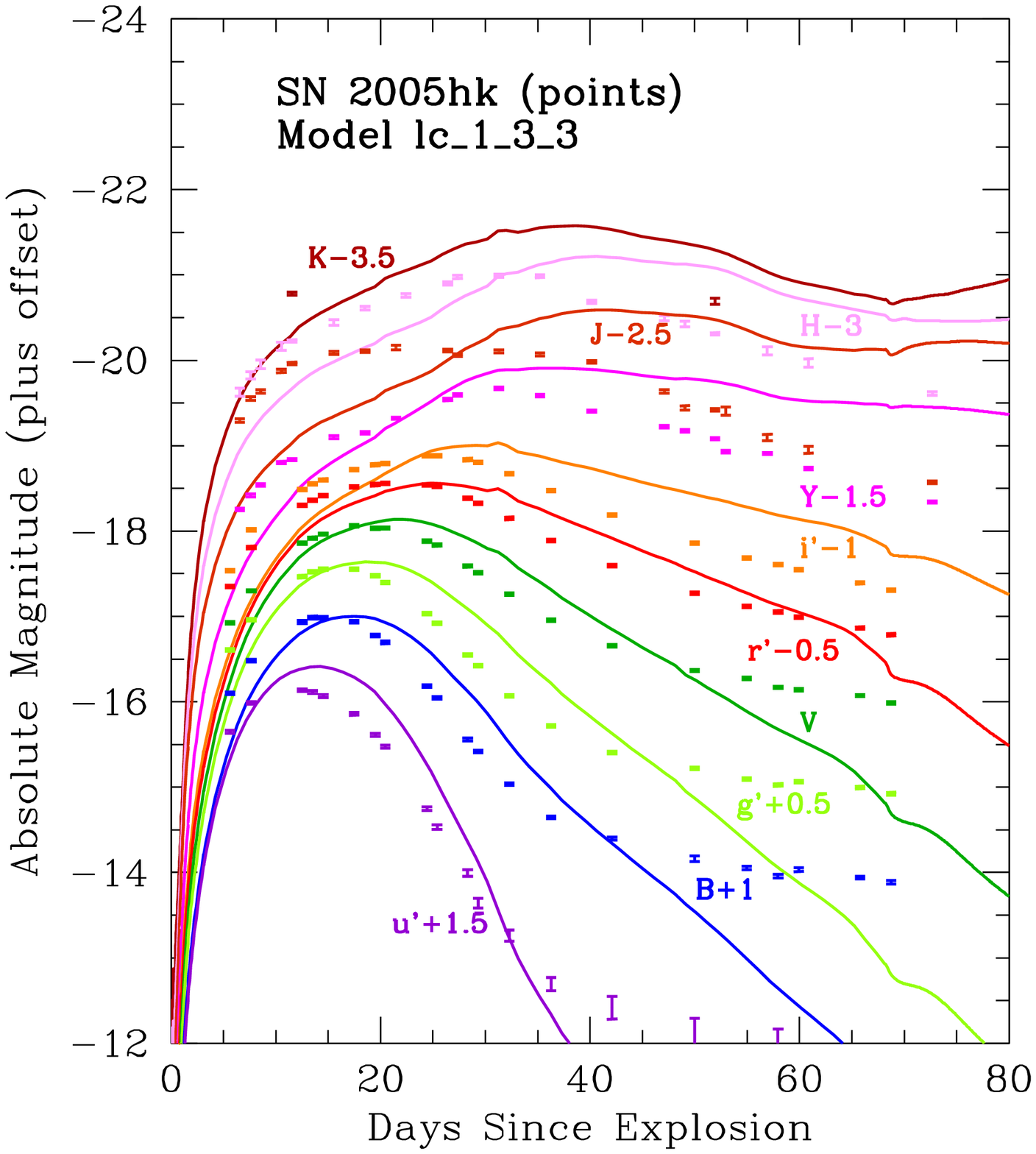]{Comparison of calculations of the $u'BVg'r'i'YJHK$
  light curves for 
  3D deflagration model \emph{1\_3\_3} with observations of SN~2005hk.
  The absolute magnitudes for SN~2005hk were calculated assuming a Galactic 
  $E(B-V)_{\mathrm {Gal}} = 0.022$ mag \citep{Sch_etal98}, a host-galaxy reddening
  of $E(B-V)_{\mathrm {Host}} = 0.09$ mag, and a Hubble constant of
  $H_0 = 72$ km s$^{-1}$ Mpc$^{-1}$.  The explosion is assumed to have occurred 
  15~days before $B$ maximum.\label{fig:modellc}} 

\clearpage
\begin{deluxetable} {lcclcc}
\tabletypesize{\scriptsize}
\tablecolumns{6}
\tablewidth{0pt}
\tablecaption{Spectroscopic Observations of SN 2005hk\label{tab:speclog}}
\tablehead{
\colhead{UT Date} &
\colhead{JD} & 
\colhead{Days since} & 
 &
\colhead{Wavelength} & 
\colhead{Resolution} \\
 &
\colhead{-2,453,000} & 
\colhead{$B_{max}$} &
\colhead{Telescope/Instrument} & 
\colhead{Range (\AA)} &
\colhead{(\AA)$^a$}}
\startdata
2005~Nov~2.2  & 676.7  &  -08  &   Lick~3.0~m/Kast     &  3334-10400  &  5/9  \\ 
2005~Nov~3.2  & 677.7  &  -07  &   MDM~2.4~m/CCDS      &  3880-7340   &   15  \\
2005~Nov~4.2  & 678.7  &  -06  &   APO~3.5~m/DIS       &  3600-9600   &    6  \\
2005~Nov~4.2  & 678.7  &  -06  &   MDM~2.4~m/CCDS      &  3880-7340   &   15  \\
2005~Nov~5.4  & 679.9  &  -05  &   Keck1/LRIS          &  3170-9240   &  6/9  \\
2005~Nov~6.3  & 680.8  &  -04  &   APO~3.5~m/DIS       &  3600-9600   &    6  \\
2005~Nov~6.3  & 680.8  &  -04  &   Keck2/DEIMOS        &  4883-9987   &    3  \\
2005~Nov~7.2  & 681.7  &  -03  &   Keck2/DEIMOS        &  4883-10000  &    3  \\
2005~Nov~14.2 & 688.7  &  +04  &   MDM~2.4~m/CCDS      &  3878-7335   &   15  \\
2005~Nov~23.2 & 697.7  &  +13  &   LCO~2.5~m/ModSpec   &  3780-7290   &    7  \\
2005~Nov~25.1 & 699.6  &  +15  &   SMARTS~1.5~m/RCSpec &  3200-9560   &   14   \\
2005~Dec~4.3  & 708.8  &  +24  &   Keck1/LRIS          &  3280-9320   & 6/12  \\
2005~Dec~7.3  & 711.8  &  +27  &   MDM~2.4~m/CCDS      &  3810-7267   &   15  \\
2005~Dec~18.1 & 722.6  &  +38  &   NTT/EMMI            &  4000-10200  &    8  \\
2005~Dec~20.1 & 724.6  &  +40  &   LCO~2.5~m/WFCCD     &  3800-9235   &    6  \\
2005~Dec~23.1 & 727.6  &  +43  &   LCO~2.5~m/WFCCD     &  3800-9235   &    6  \\
2005~Dec~24.1 & 728.6  &  +44  &   LCO~2.5~m/WFCCD     &  3800-9235   &    6  \\
2006~Jan~4.1  & 739.6  &  +55  &   MDM~2.4~m/CCDS      &  3915-7370   &   15  \\
2006~Jan~16.0 & 751.5  &  +67  &   NTT/EMMI            &  4000-10200  &    8  \\
\enddata
\tablenotetext{a} {When two numbers are given for the wavelength resolution,
they correspond to the blue and red halves of a double spectrograph.}
\end{deluxetable}

\clearpage
\begin{deluxetable} {lcccc}
\tabletypesize{\scriptsize}
\tablecolumns{5}
\tablewidth{0pt}
\tablecaption{CSP $u'g'r'i'$ Photometry of Comparison Stars in the Field of SN 2005hk\label{tab:stds1}}
\tablehead{
\colhead{Star$^a$} & 
\colhead{$u'$} &
\colhead{$g'$} &
\colhead{$r'$} &
\colhead{$i'$}}
\startdata
s04 & 15.591(011) & 14.453(011) & 14.502(009) & 14.580(009) \\
s05 & 16.701(013) & 15.008(011) & 14.411(015) & 14.186(015) \\
s06 & 18.090(024) & 16.436(007) & 15.811(007) & 15.575(007) \\
s07 & 19.818(069) & 17.056(007) & 15.845(007) & 15.344(008) \\
s08 & 17.954(019) & 16.496(007) & 15.946(007) & 15.719(007) \\
s09 & 18.723(039) & 16.700(007) & 15.908(007) & 15.571(008) \\
s10 & 17.284(015) & 16.247(007) & 16.058(007) & 16.026(009) \\
s11 & 17.746(014) & 16.687(007) & 16.317(007) & 16.167(008) \\
s12 & 20.154(081) & 17.573(010) & 16.484(007) & 16.047(008) \\
s13 & 18.675(040) & 17.264(007) & 16.679(007) & 16.441(007) \\
s14 & 19.854(058) & 17.660(008) & 16.807(007) & 16.490(008) \\
s15 & 20.057(067) & 17.841(008) & 17.000(007) & 16.684(007) \\
s16 & 19.191(033) & 17.628(009) & 17.006(007) & 16.741(007) \\
s17 & 18.486(019) & 17.409(015) & 17.046(007) & 16.919(009) \\
s18 & 19.572(074) & 17.841(008) & 17.190(007) & 16.950(010) \\
s19 & 21.189(293) & 18.737(009) & 17.381(012) & 16.577(011) \\
s20 & 20.825(148) & 18.546(008) & 17.693(014) & 17.330(015) \\
s21 & 21.864(547) & 18.919(013) & 17.544(007) & 16.124(012) \\
s22 & 21.250(311) & 19.145(025) & 17.840(008) & 17.235(010) \\
s23 & 19.455(065) & 18.488(009) & 18.176(012) & 18.064(013) \\
s24 & 19.727(111) & 18.900(009) & 18.543(020) & 18.365(012) \\
s25 & 20.093(169) & 19.207(013) & 18.867(013) & 18.748(020) \\
s26 & 22.835(300) & 20.028(029) & 18.745(011) & 17.896(018) \\
s27 & 22.503(851) & 20.073(027) & 18.800(011) & 17.866(017) \\
s28 & 19.911(317) & 20.447(046) & 19.197(016) & 18.218(016)  \\
\enddata
\tablenotetext{a} {The identifications correspond to those
in Figure~\ref{fig:fc}.}
\tablecomments{Uncertainties given in parentheses in thousandths of a
  magnitude correspond to the root-mean square (rms) of the magnitudes 
  obtained on five photometric nights.}
\end{deluxetable}

\clearpage
\begin{deluxetable} {lcccccccccccc}
\tabletypesize{\tiny}
\rotate
\tablecolumns{13}
\tablewidth{0pt}
\tablecaption{$UBVRI$ Photometry of Comparison Stars in the Field of SN 2005hk\label{tab:stds2}}
\tablehead{
\colhead{Star$^a$} & 
\colhead{$U$} & \colhead{$U$} &
\colhead{$B$} & \colhead{$B$} & \colhead{$B$} &
\colhead{$V$} & \colhead{$V$} & \colhead{$V$} &
\colhead{$R$} & \colhead{$R$} &
\colhead{$I$} & \colhead{$I$} \\
\colhead{ID} &
\colhead{KAIT} & \colhead{CTIO} &
\colhead{CSP} & \colhead{KAIT} & \colhead{CTIO} &
\colhead{CSP} & \colhead{KAIT} & \colhead{CTIO} &
\colhead{KAIT} & \colhead{CTIO} &
\colhead{KAIT} & \colhead{CTIO}}
\startdata
s01 & 14.516(020) & 14.508(043) &   \nodata   & 14.447(020) & 14.403(021) &   \nodata   & 13.766(005) & 13.740(016) & 13.387(016) & 13.351(018) & 13.001(003) & 12.983(025) \\
s02 &   \nodata   &   \nodata   &   \nodata   & 15.758(020) &   \nodata   &   \nodata   & 14.742(020) &   \nodata   & 14.193(020) &   \nodata   & 13.709(020) &   \nodata   \\
s03 &   \nodata   & 15.252(043) &   \nodata   &   \nodata   & 14.969(021) &   \nodata   &   \nodata   & 14.227(015) &   \nodata   & 13.802(019) &   \nodata   & 13.406(025) \\
s04 & 14.644(020) & 14.628(043) & 14.605(008) & 14.635(016) & 14.614(021) & 14.448(007) & 14.472(014) & 14.461(015) & 14.392(018) & 14.350(018) & 14.244(004) & 14.216(025) \\
s05 & 15.875(020) & 15.877(046) & 15.454(008) & 15.486(027) & 15.459(021) & 14.652(014) & 14.668(016) & 14.654(015) & 14.228(028) & 14.194(021) & 13.815(019) & 13.789(027) \\
s06 & 17.193(020) & 17.201(046) & 16.893(010) & 16.916(026) & 16.882(021) & 16.062(006) & 16.084(015) & 16.069(015) & 15.633(031) & 15.593(019) & 15.188(023) & 15.156(025) \\
s07 & 19.108(020) & 19.184(044) & 17.709(017) & 17.756(031) & 17.725(023) & 16.386(008) & 16.403(014) & 16.382(016) & 15.613(035) & 15.569(021) & 14.900(009) & 14.883(026) \\
s08 &   \nodata   &   \nodata   & 16.907(008) &   \nodata   &   \nodata   & 16.163(006) &   \nodata   &   \nodata   &   \nodata   &   \nodata   &   \nodata   &   \nodata   \\
s09 &   \nodata   & 17.796(086) & 17.243(013) &   \nodata   & 17.242(021) & 16.221(006) &   \nodata   & 16.232(015) &   \nodata   & 15.672(019) &   \nodata   & 15.132(025) \\
s10 & 16.479(020) & 16.390(044) & 16.491(007) & 16.494(015) & 16.481(022) & 16.112(006) & 16.118(012) & 16.111(016) & 15.887(030) & 15.882(018) & 15.619(058) & 15.639(025) \\
s11 &   \nodata   &   \nodata   & 17.000(007) &   \nodata   &   \nodata   & 16.448(006) &   \nodata   &   \nodata   &   \nodata   &   \nodata   &   \nodata   &   \nodata   \\
s12 &   \nodata   &   \nodata   & 18.187(009) &   \nodata   &   \nodata   & 16.974(010) &   \nodata   &   \nodata   &   \nodata   &   \nodata   &   \nodata   &   \nodata   \\
s13 &   \nodata   &   \nodata   & 17.693(009) &   \nodata   &   \nodata   & 16.914(006) &   \nodata   &   \nodata   &   \nodata   &   \nodata   &   \nodata   &   \nodata   \\
s14 & 18.955(020) & 19.022(113) & 18.196(009) & 18.193(012) & 18.224(025) & 17.192(007) & 17.167(019) & 17.182(016) & 16.538(016) & 16.559(019) & 16.007(043) & 16.052(026) \\
s15 &   \nodata   &   \nodata   & 18.348(015) &   \nodata   &   \nodata   & 17.370(008) &   \nodata   &   \nodata   &   \nodata   &   \nodata   &   \nodata   &   \nodata   \\
s16 & 18.415(020) & 18.401(045) & 18.065(009) & 18.095(018) & 18.062(021) & 17.256(007) & 17.277(010) & 17.274(016) & 16.808(003) & 16.784(019) & 16.314(012) & 16.307(026) \\
s17 &   \nodata   &   \nodata   & 17.763(009) &   \nodata   &   \nodata   & 17.191(007) &   \nodata   &   \nodata   &   \nodata   &   \nodata   &   \nodata   &   \nodata   \\
s18 & 18.659(020) & 18.664(043) & 18.289(014) & 18.282(016) & 18.298(026) & 17.458(007) & 17.420(030) & 17.448(023) & 16.948(017) & 16.965(020) & 16.493(035) & 16.524(025) \\
s19 &   \nodata   &   \nodata   & 19.462(021) &   \nodata   &   \nodata   & 17.980(013) &   \nodata   &   \nodata   &   \nodata   &   \nodata   &   \nodata   &   \nodata   \\
s20 &   \nodata   &   \nodata   & 19.071(014) & 19.031(041) & 18.979(075) & 18.072(010) & 18.040(027) & 18.065(015) & 17.407(021) & 17.434(023) & 16.866(030) & 16.904(026) \\
s21 &   \nodata   &   \nodata   & 19.737(032) &   \nodata   &   \nodata   & 18.129(018) &   \nodata   &   \nodata   &   \nodata   &   \nodata   &   \nodata   &   \nodata   \\
s22 &   \nodata   &   \nodata   & 19.787(026) &   \nodata   &   \nodata   & 18.400(011) &   \nodata   &   \nodata   &   \nodata   &   \nodata   &   \nodata   &   \nodata   \\
s23 &   \nodata   &   \nodata   & 18.758(017) &   \nodata   &   \nodata   & 18.284(009) &   \nodata   &   \nodata   &   \nodata   &   \nodata   &   \nodata   &   \nodata   \\
s24 &   \nodata   &   \nodata   & 19.201(016) &   \nodata   &   \nodata   & 18.685(011) &   \nodata   &   \nodata   &   \nodata   &   \nodata   &   \nodata   &   \nodata   \\
s25 &   \nodata   &   \nodata   & 19.495(029) &   \nodata   &   \nodata   & 19.004(015) &   \nodata   &   \nodata   &   \nodata   &   \nodata   &   \nodata   &   \nodata   \\
s26 &   \nodata   &   \nodata   & 20.650(108) &   \nodata   &   \nodata   & 19.329(020) &   \nodata   &   \nodata   &   \nodata   &   \nodata   &   \nodata   &   \nodata   \\
s27 &   \nodata   &   \nodata   & 20.756(064) &   \nodata   &   \nodata   & 19.370(021) &   \nodata   &   \nodata   &   \nodata   &   \nodata   &   \nodata   &   \nodata   \\
s28 &   \nodata   &   \nodata   & 21.362(120) &   \nodata   &   \nodata   & 19.713(049)  &   \nodata   &   \nodata   &   \nodata   &   \nodata   &   \nodata   &   \nodata   \\
\enddata
\tablenotetext{a} {The identifications correspond to those
in Figure~\ref{fig:fc}.}
\tablecomments{Uncertainties given in parentheses in thousandths of a
  magnitude.}
\end{deluxetable}

\clearpage
\begin{deluxetable}{lcccc}
\tablewidth{0pc}
\tabletypesize{\scriptsize}
\tablecaption{Infrared Photometric Sequence near SN 2005hk\label{tab:nirst}}
\tablehead{   \colhead{Star$^a$} & \colhead{$Y$} &
\colhead{$J_s$} & \colhead{$H$} & \colhead{$K_s$} } 
\startdata
s01 & 12.716(006) & 12.461(011) & 12.131(006) &   \nodata   \\
s02 & 13.291(008) & 12.947(018) & 12.487(008) &   \nodata   \\
s03 & 13.101(006) & 12.828(011) & 12.470(006) &   \nodata   \\
s04 & 14.102(008) & 14.016(013) & 13.917(007) &   \nodata   \\
s05 & 13.491(008) & 13.206(016) & 12.827(008) &   \nodata   \\
s06 & 14.837(009) & 14.524(012) & 14.089(006) &   \nodata   \\
s07 & 14.363(007) & 13.955(011) & 13.361(006) &   \nodata   \\
s10 & 15.514(011) & 15.340(020) & 15.160(017) &   \nodata   \\
s11 & 15.578(012) & 15.316(020) & 15.008(013) &   \nodata   \\
s12 & 15.089(017) & 14.700(017) & 14.104(009) &   \nodata   \\
s13 & 15.741(012) & 15.417(012) & 15.003(011) &   \nodata   \\
s14 & 15.660(014) & 15.295(018) & 14.794(010) & 14.707(020) \\
s15 & 15.867(010) & 15.530(020) & 14.994(019) &   \nodata   \\
s16 & 15.939(009) & 15.620(016) & 15.180(013) &   \nodata   \\
s17 & 16.283(011) & 16.046(019) & 15.742(023) &   \nodata   \\
s19 & 15.402(006) & 14.960(014) & 14.371(006) &   \nodata   \\
s20 & 16.476(025) & 16.121(028) & 15.561(014) & 15.572(039) \\
s27 & 16.626(015) & 16.143(028) & 15.621(025) &   \nodata   \\
s29 & 16.495(019) & 15.978(024) & 15.443(019) & 15.368(036) \\
s30 & 16.267(016) & 15.890(016) & 15.282(018) &   \nodata   \\
s31 & 16.880(019) & 16.522(034) & 15.975(028) &   \nodata   \\
s32 & 15.516(009) & 15.019(019) & 14.449(009) &   \nodata   \\
s33 & 16.824(019) & 16.353(022) & 15.839(024) &   \nodata   \\
s34 & 15.478(012) & 15.048(033) & 14.438(028) &   \nodata   \\
s35 & 15.120(008) & 14.691(019) & 14.109(008) &   \nodata   \\
s36 & 17.396(027) & 16.944(027) & 16.454(025) & 16.435(070) \\
\enddata
\tablenotetext{a} {The identifications correspond to those
in Fig. \ref{fig:fc}.}
\tablecomments{Uncertainties given in parentheses in thousandths of a
  magnitude correspond the the rms of the magnitudes obtained on six
  photometric nights.}
\end{deluxetable}

\clearpage
\begin{deluxetable} {lcccccc}
\tabletypesize{\scriptsize}
\tablecolumns{7}
\tablewidth{0pt}
\tablecaption{CSP $u'g'r'i'BV$ Photometry for SN 2005hk\label{tab:cspmags}}
\tablehead{
\colhead{JD} &
 &
 &
 &
 &
 & \\
\colhead{$-2,453,000$} &
\colhead{$u'$} & 
\colhead{$g'$} & 
\colhead{$r'$} & 
\colhead{$i'$} & 
\colhead{$B$} & 
\colhead{$V$}}
\startdata
675.60  & 16.841(019) & 16.755(016) & 16.895(016) & 17.115(016) & 16.815(016) & 16.875(015) \\
677.65  & 16.516(019) & 16.388(017) & 16.447(017) & 16.666(017) & 16.417(016) & 16.466(015) \\
682.59  & 16.356(019) & 15.878(016) & 15.945(016) & 16.175(016) & 15.971(016) & 15.940(015) \\
683.59  & 16.371(021) & 15.821(016) & 15.878(016) & 16.106(016) & 15.917(016) & 15.885(015) \\
684.62  & 16.412(021) & 15.792(016) & 15.832(016) & 16.067(017) & 15.921(016) & 15.836(015) \\
687.60  & 16.624(021) & 15.797(017) & 15.729(016) & 15.945(017) & 15.974(016) & 15.736(016) \\
689.66  & 16.850(023) & 15.874(017) & 15.704(017) & 15.884(017) & 16.129(016) & 15.765(016) \\
690.61  & 16.992(023) & 15.938(017) & 15.684(016) & 15.869(016) & 16.210(016) & 15.764(015) \\
694.66  & 17.719(026) & 16.312(016) & 15.705(016) & 15.781(017) & 16.715(016) & 15.913(015) \\
695.64  & 17.955(028) & 16.426(016) & 15.727(016) & 15.784(016) & 16.864(016) & 15.964(015) \\
698.59  & 18.488(041) & 16.804(016) & 15.864(016) & 15.845(016) & 17.327(016) & 16.207(015) \\
699.62  & 18.836(057) & 16.937(016) & 15.917(016) & 15.875(016) & 17.480(016) & 16.288(015) \\
702.63  & 19.230(066) & 17.284(016) & 16.098(016) & 16.006(016) & 17.851(016) & 16.543(015) \\
706.67  & 19.738(071) & 17.627(017) & 16.364(017) & 16.205(017) & 18.233(016) & 16.843(016) \\
712.57  & 20.060(128) & 17.929(016) & 16.659(016) & 16.491(016) & 18.488(018) & 17.133(015) \\
720.56  & 20.242(262) & 18.109(016) & 16.960(016) & 16.811(016) & 18.712(032) & 17.423(015) \\
725.59  & 20.492(126) & 18.247(016) & 17.122(016) & 16.984(016) & 18.818(020) & 17.524(015) \\
728.61  & 20.301(121) & 18.307(017) & 17.194(017) & 17.068(017) & 18.901(021) & 17.619(016) \\
730.63  & 20.598(127) & 18.290(016) & 17.262(016) & 17.123(016) & 18.848(021) & 17.655(015) \\
736.55  & 20.656(072) & 18.386(016) & 17.391(016) & 17.284(016) & 18.925(016) & 17.742(015) \\
739.55  & 20.592(091) & 18.444(016) & 17.469(016) & 17.366(016) & 18.993(017) & 17.808(015) \\
745.54  &  $\cdots$   & 18.486(017) & 17.584(016) & 17.498(016) & 19.062(053) & 17.903(019) \\
748.56  & 20.495(333) & 18.498(017) & 17.653(016) & 17.562(016) & 19.132(074) & 17.950(024) \\
751.55  &  $\cdots$   & 18.559(016) & 17.727(016) & 17.657(016) & 19.154(031) & 18.002(017) \\
761.53  &  $\cdots$   & 18.662(020) & 17.976(016) & 17.843(015) & 19.271(027) & 18.140(015) \\
\enddata
\tablecomments{All measurements were made on host galaxy-subtracted images.
  Uncertainties are given in parentheses in thousandths of a magnitude,
  with a minimum uncertainty of 0.015~mag for an individual measurement.}
\end{deluxetable}

\clearpage
\begin{deluxetable} {lccccc}
\tabletypesize{\scriptsize}
\tablecolumns{6}
\tablewidth{0pt}
\tablecaption{Near-Infrared Photometry for SN 2005hk\label{tab:nir}}
\tablehead{
\colhead{JD} & & & & & \\
\colhead{$-2,453,000$} & 
\colhead{$Y$} & 
\colhead{$J_s$} & 
\colhead{$H$} & 
\colhead{$K_s$} & 
\colhead{Instrument}}
\startdata
676.56 & 16.792(020) & 16.717(021) & 16.852(047) &  $\cdots$   & RetroCam \\
677.60 & 16.621(020) & 16.469(020) & 16.664(040) &  $\cdots$   & RetroCam \\
678.57 & 16.497(020) & 16.387(020) & 16.495(048) &  $\cdots$   & RetroCam \\
680.58 & 16.281(020) & 16.140(020) & 16.318(045) &  $\cdots$   & RetroCam \\
681.59 & 16.253(020) & 16.087(020) & 16.294(020) & 16.218(020) & WIRC \\
685.65 & 16.009(020) & 15.947(020) & 16.019(031) &  $\cdots$   & RetroCam \\
688.67 & 15.926(020) & 15.878(020) & 15.856(020) &  $\cdots$   & RetroCam \\
691.64 & 15.766(020) & 15.893(028) &  $\cdots$   &  $\cdots$   & RetroCam \\
692.62 &  $\cdots$   &  $\cdots$   & 15.737(020) &  $\cdots$   & RetroCam \\
696.66 & 15.531(020) & 15.847(020) & 15.605(020) &  $\cdots$   & RetroCam \\
697.60 & 15.475(020) & 15.914(020) & 15.532(020) &  $\cdots$   & RetroCam \\
701.62 & 15.415(020) & 15.871(020) & 15.526(020) &  $\cdots$   & RetroCam \\
705.58 & 15.540(020) & 15.974(020) & 15.590(020) &  $\cdots$   & RetroCam \\
710.60 & 15.641(020) & 16.060(020) & 15.751(020) &  $\cdots$   & RetroCam \\
717.61 & 15.887(020) & 16.390(020) & 16.015(031) &  $\cdots$   & RetroCam \\
719.62 & 15.912(020) & 16.524(024) & 16.078(035) &  $\cdots$   & RetroCam \\
722.54 & 16.007(020) & 16.628(020) & 16.210(020) & 16.307(035) & WIRC \\
723.55 & 16.036(020) & 16.647(048) &  $\cdots$   &  $\cdots$   & RetroCam \\
727.57 & 16.177(020) & 16.909(035) & 16.370(046) &  $\cdots$   & RetroCam \\
731.57 & 16.316(020) & 17.072(040) & 16.509(049) &  $\cdots$   & RetroCam \\
743.54 & 16.750(020) & 17.478(020) & 16.907(021) &  $\cdots$   & WIRC \\
756.57 & 17.171(020) & 17.826(032) &  $\cdots$   &  $\cdots$   & WIRC \\
\enddata
\tablecomments{All RetroCam measurements were made on host galaxy-subtracted 
  images; no host galaxy subtraction was performed on the WIRC images.
  Uncertainties are given in parentheses in thousandths of a magnitude,
  with a minimum uncertainty of 0.020~mag for an individual measurement.}
\end{deluxetable}

\clearpage

\begin{deluxetable} {lcccccl}
\tabletypesize{\scriptsize}
\tablecolumns{7}
\tablewidth{0pt}
\tablecaption{KAIT and CTIO $UBVRI$ Photometry for SN 2005hk\label{tab:kaitmags}}
\tablehead{
\colhead{JD} &
 &
 &
 &
 &
 & \\
\colhead{$-2,453,000$} &
\colhead{$U$} & 
\colhead{$B$} & 
\colhead{$V$} & 
\colhead{$R$} & 
\colhead{$I$} &
\colhead{Source}}
\startdata
675.59  & 16.049(009) & 16.813(006) & 16.870(009) & 16.762(011) & 16.754(020) & CTIO \\
675.73  &   \nodata   & 16.766(016) & 16.821(015) & 16.654(015) & 16.635(019) & KAIT \\
677.73  &   \nodata   & 16.355(016) & 16.420(022) & 16.246(020) & 16.208(021) & KAIT \\
679.78  &   \nodata   & 16.129(015) & 16.149(015) & 15.997(015) & 15.972(031) & KAIT \\
680.70  &   \nodata   & 16.054(016) & 16.057(015) & 15.925(015) & 15.903(015) & KAIT \\
684.75  &   \nodata   & 15.926(016) & 15.830(027) & 15.627(022) &   \nodata   & KAIT \\
689.70  &   \nodata   & 16.089(032) & 15.729(015) & 15.525(015) & 15.453(097) & KAIT \\
690.75  &   \nodata   & 16.202(016) & 15.748(015) & 15.555(015) & 15.472(016) & KAIT \\
691.71  &   \nodata   & 16.308(016) & 15.762(015) & 15.538(015) & 15.435(015) & KAIT \\
692.70  &   \nodata   & 16.443(015) & 15.787(015) & 15.542(015) & 15.390(015) & KAIT \\
693.67  &   \nodata   & 16.561(015) & 15.803(016) & 15.527(015) & 15.388(015) & KAIT \\
694.72  &   \nodata   & 16.710(015) & 15.857(015) & 15.544(015) & 15.346(016) & KAIT \\
695.74  &   \nodata   & 16.877(015) & 15.919(015) &   \nodata   &   \nodata   & KAIT \\
696.66  &   \nodata   & 16.982(015) & 15.973(015) & 15.602(015) & 15.383(015) & KAIT \\
697.65  &   \nodata   & 17.125(017) & 16.045(015) & 15.616(015) & 15.420(015) & KAIT \\
698.68  &   \nodata   & 17.292(015) & 16.134(015) & 15.691(015) & 15.442(015) & KAIT \\
701.55  & 18.298(060) & 17.667(014) & 16.405(006) & 15.880(006) & 15.583(008) & CTIO \\
701.71  &   \nodata   & 17.664(015) & 16.385(015) & 15.832(015) & 15.527(015) & KAIT \\
708.67  &   \nodata   & 18.171(023) & 16.841(015) & 16.225(015) & 15.833(017) & KAIT \\
710.69  &   \nodata   & 18.291(022) & 16.992(016) & 16.358(015) & 15.948(015) & KAIT \\
718.63  &   \nodata   & 18.713(040) & 17.285(017) & 16.715(015) & 16.471(063) & KAIT \\
729.66  &   \nodata   & 18.760(022) & 17.548(017) & 17.012(016) & 16.645(015) & KAIT \\
\enddata
\tablecomments{No host galaxy subtraction was performed on the images.
  Uncertainties are given in parentheses in thousandths of a magnitude.}
\end{deluxetable}

\clearpage

\begin{deluxetable} {lcccccl}
\tabletypesize{\scriptsize}
\tablecolumns{7}
\tablewidth{0pt}
\tablecaption{SDSS and MDM $ugriz$ Photometry for SN 2005hk\label{tab:sdssmags}}
\tablehead{
\colhead{JD} &
 &
 &
 &
 &
 & \\
\colhead{$-2,453,000$} &
\colhead{$u$} & 
\colhead{$g$} & 
\colhead{$r$} & 
\colhead{$i$} & 
\colhead{$z$} &
\colhead{Source}}
\startdata
669.76 &   \nodata   & $> 24.02^a$ & $> 22.68^a$ & $> 23.04^a$ & $> 21.83^a$ & SDSS~II \\
671.84 & 18.626(053) & 18.748(016) & 18.953(026) & 19.290(035) & 19.665(101) & SDSS~II \\
674.74 & 17.066(021) & 16.996(018) & 17.120(023) & 17.353(020) & 17.595(024) & SDSS~II \\
676.73 &   \nodata   & 16.526(036) & 16.651(017) & 16.855(048) & 17.012(066) & MDM \\
676.83 & 16.687(021) & 16.526(019) & 16.600(012) & 16.811(014) & 17.014(016) & SDSS~II \\
678.75 &   \nodata   & 16.120(097) & 16.281(026) & 16.529(019) & 16.696(012) & MDM \\
679.80 & 16.454(047) & 16.053(011) & 16.177(014) & 16.403(026) & 16.569(020) & SDSS~II \\
681.79 & 16.441(038) & 15.918(015) & 15.997(011) & 16.229(014) & 16.369(015) & SDSS~II \\
685.75 & 16.591(045) & 15.789(020) & 15.745(031) & 16.015(026) & 16.130(022) & SDSS~II \\
694.77 & 17.903(113) & 16.402(044) & 15.689(018) & 15.775(015) & 15.871(025) & SDSS~II \\
696.74 &   \nodata   &   \nodata   & 15.706(033) & 15.831(028) & 15.875(075) & MDM \\
697.75 & 18.494(035) & 16.781(021) & 15.825(025) & 15.819(024) & 15.931(038) & SDSS~II \\
699.78 &   \nodata   & 17.069(134) & 15.904(053) & 15.893(023) & 15.974(034) & MDM \\
700.75 & 19.047(040) & 17.156(038) & 15.986(023) & 15.902(033) & 15.990(023) & SDSS~II \\
705.73 & 19.696(051) & 17.635(010) & 16.303(010) & 16.204(015) & 16.206(017) & SDSS~II \\
730.56 &   \nodata   &   \nodata   & 17.341(007) & 17.252(009) &   \nodata   & MDM \\
739.58 &   \nodata   & 18.433(024) & 17.558(014) & 17.498(030) & 17.309(024) & MDM \\
\enddata
\tablenotetext{a} {3$\sigma$ upper limit using the asinh magnitudes adopted 
  by SDSS.}
\tablecomments{All measurements were derived from the scene modeling 
  algorithm of \citet{Hol_etal07}.
  Uncertainties are given in parentheses in thousandths of a magnitude.}
\end{deluxetable}

\clearpage
\begin{deluxetable} {lcc}
\tabletypesize{\scriptsize}
\tablecolumns{3}
\tablewidth{0pt}
\tablecaption{Apparent Peak Magnitudes for SN 2005hk\label{tab:appmags}}
\tablehead{
 &
\colhead{JD} &
\colhead{Maximum} \\
\colhead{Filter} & 
\colhead{-2,453,000} &
\colhead{Magnitude}}
\startdata
$u'$ & 681.5 $\pm$ 0.5 & 16.34 $\pm$ 0.02 \\
$g'$ & 686.0 $\pm$ 0.5 & 15.79 $\pm$ 0.01 \\
$r'$ & 691.9 $\pm$ 0.5 & 15.68 $\pm$ 0.01 \\
$i'$ & 695.0 $\pm$ 1.0 & 15.80 $\pm$ 0.02 \\
$B$  & 685.1 $\pm$ 0.5 & 15.92 $\pm$ 0.01 \\
$V$  & 688.7 $\pm$ 0.5 & 15.75 $\pm$ 0.01 \\
$R$  & 691.4 $\pm$ 0.5 & 15.53 $\pm$ 0.02 \\
$I$  & 695.1 $\pm$ 0.5 & 15.38 $\pm$ 0.02 \\
$Y$  & 701.2 $\pm$ 1.0 & 15.41 $\pm$ 0.02 \\
$J$  & 696.4 $\pm$ 8.0 & 15.88 $\pm$ 0.03 \\
$H$  & 700.9 $\pm$ 1.5 & 15.52 $\pm$ 0.02 \\
\enddata
\tablecomments{Uncertainties in the peak magnitudes were estimated
  from the rms of the photometric points about a polynomial
  fit. Uncertainties in the epochs of maximum are based on the
  light-curve sampling for each filter and the shape of the light curve.} 
\end{deluxetable}

\clearpage
\begin{deluxetable} {lcccccc}
\tabletypesize{\scriptsize}
\tablecolumns{7}
\tablewidth{0pt}
\tablecaption{Absolute Peak Magnitudes\label{tab:absmags}}
\tablehead{
\colhead{SN} &
\colhead{M$^{max}_{B}$} &
\colhead{M$^{max}_{V}$} &
\colhead{M$^{max}_{R}$} &
\colhead{M$^{max}_{I}$} &
\colhead{M$^{max}_{J}$} &
\colhead{M$^{max}_{H}$}}
\startdata
2002cx & -17.53(26) & -17.49(22) & -17.60(20) & -17.73(18) &  $\cdots$  &  $\cdots$  \\
2005hk & -18.02(32) & -18.08(29) & -18.20(27) & -18.28(26) & -17.70(25) & -18.02(25) \\
Typical SN~Ia$^a$ & -19.12(04) & -19.06(04) & -19.09(03) & -18.85(05) & -18.61(13) & -18.28(15) \\
1999by$^b$ & -17.15(23) & -17.64(23) & -17.84(23) & -17.86(23) & \nodata & -17.87(15) \\
\enddata
\tablenotetext{a}{These absolute magnitudes correspond to a normal SN~Ia with a decline rate of
$\Delta$m$_{15}$($B$) = 1.56 mag.}
\tablenotetext{b}{SN~1999by was a low-luminosity SN~1991bg-like event.}
\end{deluxetable}

\clearpage

\begin{deluxetable} {lccccl}
\tabletypesize{\scriptsize}
\tablecolumns{6}
\tablewidth{0pt}
\tablecaption{Revised Photometry of SN 2002cx\label{tab:bvri02cx}}
\tablehead{
\colhead{JD} &
 &
 &
 &
 & \\
\colhead{$-2,452,000$} &
\colhead{$B$} & 
\colhead{$V$} & 
\colhead{$R$} & 
\colhead{$I$} &
\colhead{Telescope}}
\startdata
411.78 & 17.920(035) & 17.995(018) & 17.868(017) & 17.926(020) & Nickel \\
412.78 & 17.755(021) & 17.896(027) & 17.886(032) & 17.601(042) & KAIT   \\
419.81 &   \nodata   &   \nodata   & 17.644(078) & 17.377(070) & KAIT   \\
421.76 & 18.180(058) & 17.764(043) & 17.576(038) & 17.381(055) & KAIT   \\
422.74 & 18.220(041) & 17.757(025) & 17.575(023) & 17.424(048) & KAIT   \\
423.73 & 18.339(025) & 17.861(025) & 17.564(071) & 17.492(034) & KAIT   \\
424.75 & 18.576(034) & 17.916(028) & 17.625(022) & 17.501(052) & KAIT   \\
425.75 & 18.771(048) & 17.958(030) & 17.651(032) & 17.539(054) & KAIT   \\
426.71 & 18.879(051) & 17.989(039) & 17.640(042) & 17.506(048) & KAIT   \\
427.71 & 19.047(088) & 18.095(044) & 17.681(025) & 17.510(063) & KAIT   \\
428.75 & 19.232(061) & 18.190(037) & 17.779(031) & 17.563(056) & KAIT   \\
429.74 & 19.424(068) & 18.248(042) & 17.756(030) & 17.552(043) & KAIT   \\
430.70 & 19.547(115) & 18.368(039) & 17.879(029) & 17.638(047) & KAIT   \\
431.76 & 19.543(206) & 18.440(059) & 17.858(047) & 17.509(062) & KAIT   \\
432.69 & 19.783(149) & 18.477(040) & 17.923(027) & 17.584(044) & KAIT   \\
433.73 & 19.985(031) & 18.691(025) & 18.018(019) &   \nodata   & Nickel \\
433.78 & 20.041(037) & 18.670(023) &   \nodata   & 17.632(015) & Nickel \\
433.79 &   \nodata   & 18.613(031) &   \nodata   &   \nodata   & Nickel \\
434.70 & 19.975(176) & 18.574(087) & 17.974(030) & 17.651(075) & KAIT   \\
435.69 & 19.974(146) & 18.739(047) & 18.064(035) & 17.844(078) & KAIT   \\
435.71 & 20.142(035) & 18.826(035) & 18.124(015) & 17.702(034) & Nickel \\
435.71 &   \nodata   & 18.840(035) &   \nodata   &   \nodata   & Nickel \\
436.74 & 20.275(126) & 18.789(083) & 18.172(046) & 17.777(065) & KAIT   \\
436.75 & 20.277(035) & 18.933(024) & 18.174(016) & 17.916(024) & Nickel \\
437.70 & 20.270(195) & 18.922(072) & 18.335(040) & 17.867(056) & KAIT   \\
437.70 & 20.362(041) & 18.944(032) & 18.222(015) & 17.920(032) & Nickel \\
437.74 & 20.315(031) &   \nodata   &   \nodata   &   \nodata   & KAIT   \\
438.72 & 20.222(140) & 18.952(055) & 18.485(148) & 17.843(075) & KAIT   \\
440.71 &   \nodata   & 19.153(079) & 18.381(047) & 18.058(058) & KAIT   \\
442.71 &   \nodata   & 19.155(081) & 18.525(068) & 18.088(081) & KAIT   \\
443.74 &   \nodata   & 19.290(143) & 18.675(091) & 18.216(111) & KAIT   \\
447.71 &   \nodata   & 19.396(144) & 18.833(105) & 18.261(087) & KAIT   \\
465.69 & 21.265(113) & 19.821(046) & 19.305(053) & 18.820(163) & Nickel \\
621.08 &   \nodata   & 21.598(164) & 21.546(229) & 20.751(201) & Nickel \\
\enddata
\tablecomments{All measurements were made on host galaxy-subtracted images.
  Uncertainties are given in parentheses in thousandths of a magnitude.}
\end{deluxetable}

\clearpage

\begin{figure}
\plotone{f1.eps}
{\center Phillips {\it et al.} Fig. \ref{fig:fc}}
\end{figure}

\begin{figure}
\epsscale{0.93}
\plotone{f2.eps}
{\center Phillips {\it et al.} Fig. \ref{fig:spmontage}}
\end{figure}

\begin{figure}
\epsscale{0.98}
\plotone{f3.eps}
{\center Phillips {\it et al.} Fig. \ref{fig:gBV}}
\end{figure}

\begin{figure}
\epsscale{0.92}
\plotone{f4.eps}
{\center Phillips {\it et al.} Fig. \ref{fig:csplc}}
\end{figure}

\begin{figure}
\epsscale{0.90}
\plotone{f5.eps}
{\center Phillips {\it et al.} Fig. \ref{fig:kaitsdsslc}}
\end{figure}

\begin{figure}
\epsscale{0.96}
\plotone{f6.eps}
{\center Phillips {\it et al.} Fig. \ref{fig:cspkait}}
\end{figure}

\begin{figure}
\epsscale{0.92}
\plotone{f7.eps}
{\center Phillips {\it et al.} Fig. \ref{fig:cspsdss}}
\end{figure}

\begin{figure}
\epsscale{0.94}
\plotone{f8.eps}
{\center Phillips {\it et al.} Fig. \ref{fig:02cx_05hk_spec}}
\end{figure}

\begin{figure}
\epsscale{0.94}
\plotone{f9.eps}
{\center Phillips {\it et al.} Fig. \ref{fig:02cx_05hk_vels}}
\end{figure}

\begin{figure}
\epsscale{0.97}
\plotone{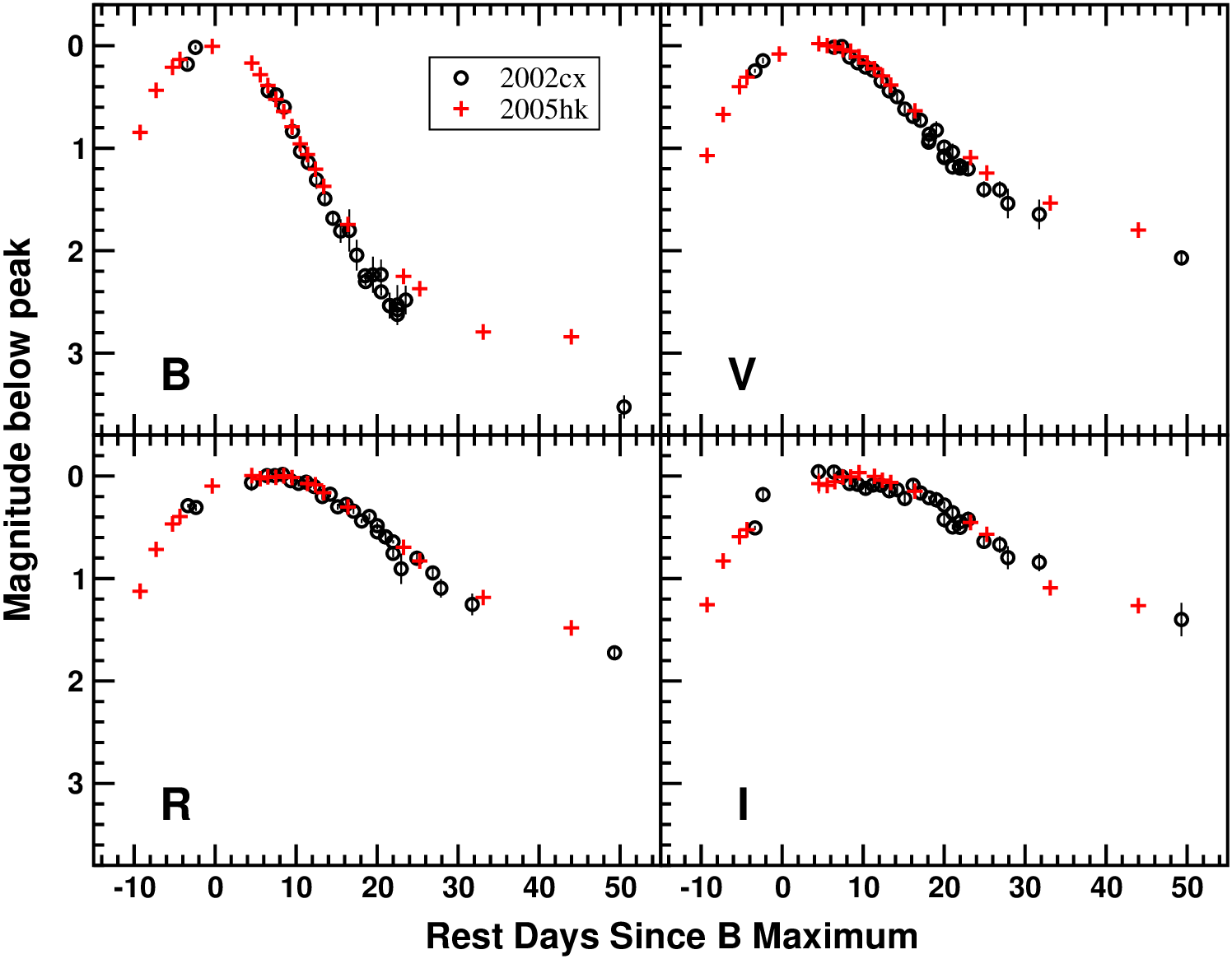}
{\center Phillips {\it et al.} Fig. \ref{fig:02cx_05hk_phot}}
\end{figure}

\begin{figure}
\epsscale{0.93}
\plotone{f11.eps}
{\center Phillips {\it et al.} Fig. \ref{fig:02cx_05hk_colors}}
\end{figure}

\begin{figure}
\epsscale{0.86}
\plotone{f12.eps}
{\center Phillips {\it et al.} Fig. \ref{fig:nir}}
\end{figure}

\begin{figure}
\epsscale{0.93}
\plotone{f13.eps}
{\center Phillips {\it et al.} Fig. \ref{fig:optir}}
\end{figure}

\begin{figure}
\epsscale{0.87}
\plotone{f14.eps}
{\center Phillips {\it et al.} Fig. \ref{fig:pabsmags}}
\end{figure}

\begin{figure}
\plotone{f15.eps}
{\center Phillips {\it et al.} Fig. \ref{fig:uvoir}}
\end{figure}

\begin{figure}
\plotone{f16.eps}
{\center Phillips {\it et al.} Fig. \ref{fig:modellc}}
\end{figure}


\begin{thebibliography}{}

\bibitem[Aldering et al.(2002)]{Ald_etal02} Aldering, G., et al. 2002,
  \procspie, 4836, 61

\bibitem[Antilogus et al.(2005)]{Ant_etal05} Antilogus, P., et al. 2005,
  The Astronomer's Telegram, 502, 1

\bibitem[{{Arnett}(1979)}]{Arn79} Arnett, W.~D.\ 1979, \apjl, 
230, L37 

\bibitem[Arnett(1982)]{Arn82} Arnett, W. D. 1982, \apj, 253, 785

\bibitem[Arnett, Branch, \& Wheeler(1985)]{Arn_etal85} Arnett, W. D., 
  Branch, D., \& Wheeler, J. C. 1985, \nat, 314, 337

\bibitem[Astier et al.(2006)]{Ast_etal06} Astier, P., et al. 2006, \aap,
  447, 31

\bibitem[Baron, Lentz, \& Hauschildt(2003)]{Bar_etal03} Baron, E., Lentz,
  E. J., \& Hauschildt, P. H. 2003, \apj, 588, L29

\bibitem[Barentine et al.(2005)]{Bar_etal05} Barentine, J., et al. 2005, CBET,
  268

\bibitem[Blinnikov et al.(2006)]{Bli_etal06} Blinnikov, S. I., et al. 2006,
  \aap, 453, 229

\bibitem[Blondin et al.(2006)]{Blo_etal06} Blondin, S., et al. 2006, \aj,
  131, 1648

\bibitem[Branch et al.(2004)]{Bra_etal04} Branch, D., Baron, E., Thomas, R. C.,
  Kasen, D., Li, W., \& Filippenko, A. V. 2004, \pasp, 116, 903

\bibitem[Branch et al.(2006)]{Bra_etal06} Branch, D., Dang, L. C., Hall, N.,
  Ketchum, W., Melakayil, M., Parrent, J., Troxel, M. A., Casebeer, D., 
  Jeffery, D. J, \& Baron, E. 2006, \pasp, 118, 560

\bibitem[Burket \& Li(2005)]{Bur05} Burket, J., \& Li, W. 2005, \iaucirc,
  8625

\bibitem[Candia et al.(2003)]{Can_etal03} Candia, P., et al. 2003, \pasp,
  115, 277

\bibitem[Cardelli, Clayton, \& Mathis(1989)]{Car_etal89} Cardelli, J. A.,
  Clayton, G. C., \& Mathis, J. S. 1989, \apj, 345, 245

\bibitem[Chornock et al.(2006)]{Cho_etal06} Chornock, R., Filippenko, A. V.,
  Branch, D., Foley, R. J., Jha, S., \& Li, W. 2006, \pasp, 118, 722

\bibitem[Conley et al.(2006)]{Con_etal06} Conley, A., et al. 2006, \aj,
  132, 1707

\bibitem[Contardo et al.(2000)]{Con00}
Contardo, G., Leibundgut, B., \& Vacca, W.~D. 2000, \aap, 359, 876
 
\bibitem[Filippenko (1997)]{Fil_97} Filippenko, A. V.,
  1997, ARAA, 35, 309
 
\bibitem[Filippenko (2005)]{Fil_05} Filippenko, A. V., in
   White Dwarfs: Cosmological and Galactic Probes,
  ed. E. M. Sion, S. Vennes, \& H. L. Shipman (Dordrecht: Springer), 97  
 
\bibitem[Filippenko et al.(1992a)]{Fil_etal92a} Filippenko, A. V.,
  et al. 1992a, ApJ, 384, L15
 
\bibitem[Filippenko et al.(1992b)]{Fil_etal92b} Filippenko, A. V.,
  et al. 1992b, AJ, 104, 1543
 
\bibitem[Filippenko et al.(2001)]{Fil_etal01} Filippenko, A. V.,
  Li, W. D., Treffers, R. R., \& Modjaz, M. 2001, in Small 
  Telescope Astronomy on Global Scales, ed. W.-P. Chen, C. Lemme, \&
  B. Paczy\'{n}ski (San Francisco: ASP), 121

\bibitem[Fisher (2000)]{Fisher00} Fisher,~A. 2000, Ph.D. thesis, 
  Univ. Oklahoma

\bibitem[Ford et al.(1993)]{For_etal93} Ford, C. H., Herbst, W., Richmond, 
  M. W., Baker, M. L., Filippenko, A. V., Treffers, R. R., Paik, Y., \& 
  Benson, P. J. 1993, \aj, 106, 1101

\bibitem[Freedman et al.(2001)]{Fre_etal01} Freedman, W. L., et al. 
   2001, \apj, 553, 47

\bibitem[Frieman et al.(2007)]{Fri_etal07} Frieman, J., et al. 2007,
  in preparation

\bibitem[Fukugita et al.(1996)]{Fuk_etal96} Fukugita, M., Ichikawa, T.,
  Gunn, J. E., Doi, M., Shimasaku, K., \& Schneider, D. P. 1996, \aj, 111,
  1748

\bibitem[Gallagher et al.(2005)]{Gal_etal05} Gallagher, J. S., Garnavich,
  P. M., Berlind, P., Challis, P., Jha, S., \& Kirshner, R. P. 2005, \apj,
  634, 210

\bibitem[Gamezo, Khokhlov, \& Oran(2004)]{Gam_etal04} Gamezo, V. N., 
  Khokhlov, A. M., \& Oran, E. S. 2004, \prl, 92, 211102

\bibitem[Ganeshalingam et al.(2007)]{gan07} Ganeshalingam, M., et al. 2007, 
  in preparation

\bibitem[Garnavich et al.(2004)]{Gar_etal04} Garnavich, P.~M., et al. 2004,
  \apj, 613, 1120

\bibitem[Goldhaber et al.(2001)]{Gol_etal01} Goldhaber, G., et al. 2001, \apj,
  558, 359

\bibitem[Gunn et al.(1998)]{Gunn_etal98} Gunn, J. E., et al. 1998,
  \aj, 116, 3040

\bibitem[Gunn et al.(2006)]{Gunn_etal06} Gunn, J. E., et al. 2006,
  \aj, 131, 2332

\bibitem[Guy et al.(2005)]{Guy_etal05} Guy, J, Astier, P., Nobili, S.,
  Regnault, N., \& Pain, R. 2005, \aap, 443, 781

\bibitem[Hamuy et al.(1995)]{Ham_etal95} Hamuy, M., Phillips, M. M., 
  Maza, J., Suntzeff, N. B., Schommer, R. A., \& Avil\'{e}s, R. 1995, \aj,
  109, 1

\bibitem[Hamuy et al.(1996)]{Ham_etal96} Hamuy, M., Phillips, M. M.,
  Suntzeff, N. B., Schommer, R. A., Maza, J., \& Avil\'{e}s, R. 1996, \aj, 
  112, 2398

\bibitem[Hamuy et al.(2006)]{Ham06} Hamuy,~M., et al. 2006, \pasp,
  118, 2

\bibitem[Hernandez et al.(2000)]{Her_etal00} Hernandez, M., et al. 2000,
  \mnras, 319, 223

\bibitem[H\"{o}flich et al.(2003)]{Hof_etal03} H\"{o}flich, P., Gerardy,
  C., Linder, E., \& Marion, H. 2003, in Stellar Candles for the 
  Extragalactic Distance Scale, ed. D.~Alloin \& W.~Gieren (Berlin: Springer),
  203

\bibitem[Holtzman et al.(2007)]{Hol_etal07} Holtzman, J., et al. 2007, in
  preparation

\bibitem[Howell(2001)]{How01} Howell, D. A., \apj, 554, L193

\bibitem[Hoyle \& Fowler(1960)]{Hoyle60} Hoyle, F., \& Fowler, W. A. 1960,
  \apj, 132, 565

\bibitem[Ivezic et al.(2007)]{Ive_etal07} Ivezic, Z., et al. 2007, in 
  preparation

\bibitem[Jha(2002)]{Jha02} Jha, S. 2002, Ph.D. thesis, Harvard Univ.

\bibitem[Jha et al.(2006)]{Jha_etal06} Jha, S., et al. 2006, \aj, 132, 189

\bibitem[Kasen(2006)]{Kasen06} Kasen, D. 2006, \apj, 649, 939

\bibitem[Kasen et al.(2004)]{Kas_etal04} Kasen, D., Nugent, P., Thomas, R. C.,
  \& Wang, L. 2004, \apj, 610, 876

\bibitem[Kozma et al.(2005)]{Koz_etal05} Kozma, C., et al. 2005, \aap, 437, 983

\bibitem[Krisciunas, Phillips, \& Suntzeff(2004)]{Kri_etal04a} Krisciunas,
  K., Phillips, M. M., \& Suntzeff, N. B. 2004, \apjl, 602, L81

\bibitem[Krisciunas et al.(2004)]{Kri_etal04b} Krisciunas, K., et al. 2004,
  \aj, 128, 3034

\bibitem[Landolt(1992)]{lan92} Landolt, A. U. 1992, \aj, 104, 340

\bibitem[Leibundgut et al.(1993)]{Leib_etal93} Leibundgut, B., et al.
   1993, \aj, 105, 301

\bibitem[Li et al.(2000)]{Li_etal00} Li, W., et al. 2000, in Cosmic 
Explosions, ed. S. S. Holt \& W. W. Zhang (New York: AIP), 103

\bibitem[Li, Filippenko, \& Riess(2001)]{Li_etal01} Li, W., Filippenko, A.~V.,
  \& Riess, A.~G. 2001, \apj, 546, 719

\bibitem[Li et al.(2003)]{Li_etal03} Li, W., et al. 2003, \pasp, 115, 453
  (LFC)

\bibitem[Lupton, Gunn, \& Szalay(1999)]{Lup_etal99} Lupton, R.~H., Gunn, J.~E.,
  \& Szalay, A.~S. 1999, \aj, 118, 1406

\bibitem[Macri et al.(2001)]{Mac_etal01} Macri, L.~M., et al. 2001, \apj,
  559, 243

\bibitem[Milne et al.(2007)]{Mil_etal07} Milne, P., et al. 2007, 
  in preparation

\bibitem[Morgan et al.(2005)]{Mor_etal05} Morgan, C. W., Byard, P. L., 
  DePoy, D. L., Derwent, M., Kochanek, C. S., Marshall, J. L., 
  O'Brien, T. P., Pogge, R. W. 2005, \aj, 129, 2504

\bibitem[Nugent et al.(1995)]{Nug_etal95} Nugent, P., Phillips, M., Baron, 
  E., Branch, D., Hauschildt, P. 1995, \apj, 455, 147

\bibitem[Perlmutter et al.(1999)]{Per_etal99} Perlmutter, S., 
   et al. 1999, \apj, 517, 565

\bibitem[Persson et~al.(1998)]{persson98} Persson, S.~E., Murphy, D.~C., 
Krzeminski, W., Roth, M., \& Rieke, M.~J. 1998, \aj, 116, 2475

\bibitem[Persson et al.(2002)]{Per_etal02} Persson, S. E., et al. 2002,
  \aj, 124, 619

\bibitem[Phillips(1993)]{Phi93} Phillips, M. M. 1993, \apjl, 413, L105

\bibitem[Phillips et al.(1992)]{Phi_etal92} Phillips, M. M.,
  et al. 1992, 103, 1632

\bibitem[Phillips et al.(1999)]{Phi_etal99} Phillips, M. M., Lira, P.,
  Suntzeff, N. B., Schommer, R. A., Hamuy, M., \& Maza, J. 1999, \aj, 118, 1766

\bibitem[Prieto, Rest, \& Suntzeff(2006)]{Pri_etal06} Prieto, J. L., Rest, A.,
  \& Suntzeff, N. B. 2006, \apj, 647, 501

\bibitem[Riess, Press, \& Kirshner(1996)]{Rie_etal96} Riess, A. G.,
Press, W. H., \& Kirshner, R. P. 1996, \apj, 473, 88

\bibitem[Riess et al.(1998)]{Rie_etal98} Riess, A. G., et al. 
   1998, \aj, 116, 1009

\bibitem[Riess et al.(1999)]{Rie_etal99} Riess, A. G., et al. 1999, \aj,
  118, 2675

\bibitem[R{\"o}pke et al.(2006)]{roepke2006} R{\"o}pke, F.~K., 
Gieseler, M., Reinecke, M., Travaglio, C., \& Hillebrandt, W.\ 2006, \aap, 
453, 203 

\bibitem[R{\"o}pke et al.(2007)]{roepke2007} R{\"o}pke, F.~K., 
 Woosley, S.E.,\& Hillebrandt, W.\ 2007, ApJ, submitted, astro-ph/0609088 

\bibitem[Schlegel, Finkbeiner, \& Davis(1998)]{Sch_etal98} Schlegel, D. J., 
  Finkbeiner, D. P., \& Davis, M. 1998, \apj, 500, 525

\bibitem[Serduke et al.(2005)]{Ser_etal05} Serduke, F. J. D., Wong, D. S., \&
  Filippenko, A. V. 2005, CBET, 269

\bibitem[Smith et al.(2002)]{Smi_etal02} Smith, J. A., et al. 2002, \aj, 123, 2121

\bibitem[Stritzinger et al.(2002)]{Str_etal02} Stritzinger, M., et al. 2002,
  \aj, 124, 2100

\bibitem[{{Stritzinger} \& {Leibundgut}(2005)}]{StrLeib05} Stritzinger, 
M., \& Leibundgut, B.\ 2005, \aap, 431, 423 

\bibitem[Stritzinger et al.(2006)]{Str_etal06} Stritzinger, M., Leibundgut, B.,
  Walch, S., \& Contardo, G. 2006, \aap, 450, 241

\bibitem[Stritzinger et al.(2007)]{Stritz07} Stritzinger, 
M., Mazzali, P.,  Sollerman, J. \& Benetti, S.\ 2007, \aap, submitted, 
astro-ph/0609232

\bibitem[Suntzeff et al.(1999)]{Sun_etal99} Suntzeff, N.~B., et al. 1999,
  \aj, 319, 1175

\bibitem[Suntzeff(2000)]{Sun00} Suntzeff, N.~B. 2000, in 
  Cosmic Explosions, ed. S.~S.~Holt \& W.~W.~Zhang (New York: AIP), 65

\bibitem[Thomas et al.(2002)]{Tho_etal02} Thomas, R. C., et al. 2002, \apj,
  567, 1037

\bibitem[Travaglio et al.(2004)]{Tra_etal04} Travaglio, C., Hillebrandt, W.,
  Reinecke, M., \& Thielemann, F.-K. 2004, \aap, 425, 1029

\bibitem[Woosley(1988)]{Woosley88}  Woosley, S. E. 1988, \apj, 330,
  218

\bibitem[{{Woosley} et al.(2007)}]{Woosley07} Woosley, S. E.,
Kasen, D., Blinnikov, S., \& Sorokina, E. \ 2007, ApJ, submitted,  
astro-ph/0609562 

\bibitem[York et al.(2000)]{Yor_etal00} York, D. G., et al. 2000, \aj, 120,
  1579



\end{thebibliography}
\end{document}